\documentclass[prb,twocolumn,amsmath,aps,amssymb,english]{revtex4-1}

\usepackage{natbib}
\usepackage{subfigure}
\usepackage{tabularx}
\usepackage{epsfig}
\usepackage{longtable}
\usepackage{amsfonts}
\usepackage{rotating}
\usepackage{subfigure}
\usepackage{hyperref}
\usepackage{babel}

\usepackage{currfile}

\usepackage{color}

\newcommand{\bpm}{\begin{pmatrix}}
\newcommand{\epm}{\end{pmatrix}}
\newcommand{\bmm}{\begin{matrix}}
\newcommand{\emm}{\end{matrix}}

\newcommand{\bea}{\begin{eqnarray}}
\newcommand{\eea}{\end{eqnarray}}
\newcommand{\la}{\label}
\newcommand{\be}{\begin{equation}}
\newcommand{\ee}{\end{equation}}



\begin{document}

\title{Constructing a Weyl semimetal by stacking one dimensional topological phases}
 
\author{Sriram Ganeshan}
\author{S. Das Sarma}
\affiliation{Condensed Matter Theory Center and Joint Quantum Institute, Department of Physics, University of Maryland, College Park, MD 20742, USA}

\date{\today}

%%%%%%%%%%%%%
\begin{abstract}

%Recent optical experiments have unravelled a novel topological classification of quasicrystals. Theoretically, this topological feature appears as a function of previously known but often discarded phase term in the Aubry-Andre (AAH) model also known as Harper model.  In this work we identify a different class of topological edge modes in the half flux state of the off-diagonal Harper model. This novel topology is shown to be connected to the Majorana physics. We will demonstrate simple experimental settings that can realize this interesting feature in a bosonic framework.
Topological semimetals in three-dimensions (e.g. Weyl semimetal) can be built by stacking two dimensional topological phases. The interesting aspect of such a construction is that even though the topological building blocks in the low dimension may be gapped, the higher dimensional semimetallic phase emerges as a gapless critical point of a topological phase transition between two distinct insulating phases. In this work, we extend this idea by constructing three-dimensional topological semimetallic phases akin to Weyl systems by stacking one-dimensional Aubry-Andre-Harper (AAH) lattice tight binding models with non-trivial topology. The generalized AAH model is a family of one dimensional tight binding models with cosine modulations in both hopping  and onsite energy terms.  In this paper, we present a two-parameter generalization of the AAH model that can access topological phases in three dimensions within a unified framework. We show that the $\pi$-flux state of this two-parameter AAH model manifests three dimensional topological semimetallic phases where the topological features are embedded in one dimension. The topological nature of the band touching points of the semimetallic phase in 3D is explicitly established both analytically and numerically from the 1D perspective.  This dimensional reduction provides a simple protocol to experimentally construct the three dimensional Brillouin zone of the topological semimetallic phases using `legos' of simple 1D double well optical lattices. We also propose Zak phase imaging of optical lattices as a tool to capture the topological nature of the band touching points. Our work provides a theoretical connection between the commensurate AAH model in 1D and Weyl semimetals in 3D, and points toward practical methods for the laboratory realization of such three dimensional topological systems in atomic optical  lattices.
 
 %\setcounter{tocdepth}{1}
%\tableofcontents

\end{abstract}
%%%%%%%%%%%%%

\maketitle

%\tableofcontents

\section{Introduction}

%A simple model capturing the Integer Quantum Hall topology is a 2D lattice with a magnetic flux which is also known as the Hofstadter model\cite{hofstadter}.  The 2D Hofstadter model can be mapped on to a 1D tight-binding model with onsite Cosine modulation known as the Aubry-Andre-Harper (AAH) model\cite{harper55,AA,note}. Symmetries play important roles in realizing topological phases beyond the well known Integer Quantum Hall Effect (IQHE). Topological phases of noninteracting Fermions in different dimensions have been classified according to the underlying symmetries\cite{ryu09}. 

Topological semimetals (TS) are gapless phases of matter where the metallic nature is attributed to an underlying band topology \cite{iridate11}. For example, in three dimensions, a broken time reversal or inversion symmetry can lead to non-degenerate valence and conduction bands to touch at some diabolic points in the Brillouin zone. These band touching points are topological in nature and are dubbed as Weyl semimetals (WSM)\cite{iridate11,balents11,hosur12,bernevig12}. The topological robustness against symmetry breaking terms of these band touching points manifests in the underlying low energy excitation structure. Expanding the Hamiltonian around these band touching points yields the 3D Weyl equation\cite{delplace12,jiang13}. The characteristic of a 3D Weyl equation is that it exhausts all the available Pauli matrices such that one cannot add an anti-commuting Pauli matrix that can generate a `mass' term or in other words open a gap at these band touching points. These low energy excitations not only allow for exotic surface effects such as fermi arcs, but also lead to non-trivial transport phenomena such as chiral magnetic effect\cite{aji12,burkov12,Kharzeev14}.

The gapless nodal region in the TS can be viewed as a critical point of a phase transition in the momentum space between two topologically distinct insulating phases residing in a lower dimensional subspace. These insulating phases are classified by the symmetry and the dimensionality of the lower dimensional manifold.  For example, a 3D WSM can be viewed as a momentum space phase transition between a Chern insulator and a normal insulator in 2D with the gapless (nodal) points appearing at the critical point of this topological phase transition. This effect has been proposed to exist in a ferromagnetic  compound $HgCr_2Se_4$ \cite{chernsemimetal}.

A natural question then arises whether one can construct TS by stacking topological phases in one dimension. So far this idea\cite{turner13,taylor14}  has not been quantified in terms of a rigorous model. The central goal of our work is to understand TS in 3D from the point of view of 1D models manifesting topological phases.  In this work, we show that the $\pi$-flux state of the generalized Aubry-Andre-Harper (AAH) model\cite{harper55,AA,note} in 1D can be used as a theoretically unifying framework to understand semimetallic phases in both two and three dimensions.  The simplest AAH model is a 1D tight-binding model with onsite cosine modulation with a phase parameter corresponding to the momentum in the second dimension\cite{hofstadter}. In the absence of discrete symmetries, 1D AAH models are topologically trivial\cite{ryu09}. Thus it is imperative that the 1D building blocks must have discrete symmetries in order to have a non-trivial topological index.  To allow for discrete symmetries within the AAH framework, this model can be expanded to a generalized AAH model that contains cosine modulations in both hopping (off-diagonal AAH) and onsite terms\cite{hiramoto89, han94, kraus2}. The off-diagonal AAH model for $\pi$-flux state has some surprising connections to the well-known Su-Schreiffer-Heeger (SSH) model\cite{ssh79} with inversion symmetry and was recently\cite{sriramaah, li} mapped to the double Majorana chain which is known to manifest a $Z_2$ topological invariant\cite{kitaev01}. This 1D topological phase was attributed to an underlying chiral symmetry which leads to topological zero energy modes in agreement with the classification scheme\cite{ryu09}. Recent work\cite{langprl12,chen14} has identified various 2D topological phases such as Haldane model within the generalized AAH scheme. AAH models with $\pi$-flux can be thought of as a 2D TS  as a function of the phase parameter of the hopping modulation in the corresponding 1D model.   In the case of the off-diagonal AAH model with $\pi$-flux, the gap closing point is a topological quantum phase transition point at which the topological zero modes terminate\cite{sriramaah}. 

To access 3D physics, we extend the generalized AAH model to contain two cosine modulations on both hopping and onsite terms. These modulations contain different phase parameters corresponding to the momentum in each additional dimension. Note that this model is a 1D real space model (with two distinct phase parameters) manifesting topological features in accordance with the symmetries present\cite{sriramaah} in 1D.   The presence of the phase parameters in the cosine terms of the AAH model provides the flexibility necessary for studying 3D topological features using an effective 1D model. Within the generalized AAH scheme, we uncover a remarkable connection between the symmetry protected gapless topological phases in one dimension and semimetallic phases in two and three dimensions. The idea is to construct the Brillouin zone (BZ) (momentum space) of the semimetallic phases in higher dimensions by stacking the real space AAH models in 1D with the phase parameters sweeping the whole BZ of higher dimensions.  Although the topological nature of the band touching points arises from the symmetries that are implicit in the 1D AAH model, the unique topological phases exclusive to higher dimensions emerge on breaking these symmetries. In this context, the AAH framework presents itself as an ideal platform to explore semimetallic phases in three dimensions with their topological aspects  rooted in 1D systems. We explicitly show the topological nature of the semimetallic phase by mapping these  one dimensional slices containing  the zero energy end states to double Majorana chains manifesting the $Z_2$ index. These zero energy modes are the topological end states of a 1D system and construct the 3D surface states in the form of Fermi arcs. Within our construction these arc states form the zero energy Fermi surface which can be attributed to the presence of the chiral symmetry in the 1D model. Our theory allows us to trace the exotic Fermi arcs or nodal lines unique to the 3D WSM. Understanding WSM with nodal points within 1D framework is advantageous in the context of anomalous transport and has been addressed recently in Ref. ~ (\onlinecite{taylor14}). Similar construction has been carried out for  3D topological insulators using Shockley model~\cite{shockley}. Our construction of Fermi arcs can be thought of as a generalization of the Shockley criterion for surface states of topological insulators to the case of surface arc states of  Weyl semimetals.

A particular advantage in working within the AAH framework is that all the parameters of this model can be accessed within the existing technology of cold atomic experiments. In the past few years, cold atom experiments have developed great control of 1D optical lattices with double well potentials\cite{porto06,porto07,billy08,roati08,bloch}. The parameters associated with the AAH models are equivalent to different double well potentials corresponding to different dimer configurations. A recent experiment (Ref.~(\onlinecite{bloch})) has been successful in imaging the Zak phase (Polarization)  between two different SSH dimer realizations of double well potentials. In the light of recent experimental advances, the 2-parameter generalized AAH model provides a protocol to construct three dimensional semimetallic phases in the optical lattice systems. We show that the spectral features of different realizations of  double well lattices form the three dimensional BZ (momentum space) of WSM on breaking the inversion symmetry. Our work provides simple methods to tune these double well potentials where each configuration is equivalent to a 1D BZ slice of a 3D WSM.  These different configurations can be constructed via experimentally tunable parameters (onsite energy and hopping modulations) of a double well optical lattice. 

This paper is organized as follows. In Sec.~\ref{model}, we define AAH models with two phase parameters with both diagonal and off-diagonal hopping elements. In Sec.~\ref{cssm}, we present the $\pi$-flux AAH model with two phase parameters and show that it can be mapped on to the double Majorana chains that manifest topological zero energy modes. In Sec.~(\ref{weyl}), we add onsite modulations and write down analytical conditions for the existence of the `topologically unavoidable' band touching points.  We show that these conditions manifest nodal lines and Weyl arcs which are the hallmark of the 3D TS phases. In Sec.~(\ref{inv}), we present an alternative way to identify the topological nature of theses band touching points using polarization (Zak phase) method to make contact with the experimental capabilities. Finally, in Sec.~\ref{ex} we discuss the experimental feasibility of our work and in Sec.~(\ref{conclude}) conclude the paper. In App.~(\ref{appendix}), we derive the two parameter AAH model starting from a 3D cubic lattice with a tilted flux in $yz$ plane.

\section{Model}\la{model}
In order to build TS in 2D and 3D, we consider a 1D model with two phase parameters that can sweep 3D BZ.  This 1D model can be derived from a parent 3D model shown in Appendix.~\ref{appendix}. We would like to emphasize that the topological properties we consider will be associated with the 1D model for any choice of the two phase parameters. This allows us to make contact with the physics of the 3D systems. We start with the following 1D tight binding model Hamiltonian,
 \begin{eqnarray}
&&H(\phi_y,\phi_z)=\nonumber\\
&&\sum^{N-1}_{n=1}t(1+\lambda_{xy}g_n(b_z,\phi^{od}_y)+\lambda_{xz}g_n(b_y,\phi^{od}_z))c_{n+1}^{\dagger}c_{n}+H.c.\nonumber\\
		&+&	\sum^{N}_{n=1}(\lambda_{y}g_n(b_z,\phi^{d}_y)+\lambda_{z}g_n(b_y,\phi^{d}_z)) c_{n}^{\dagger}c_{n}.
\la{aah1dm}
\end{eqnarray}
We have defined $ g_n(b,\phi)=\cos(2n\pi b+\phi)$. The above 1D chain in Eq.~(\ref{aah1dm}) has $N$ sites ($n=1$, $2$, $\ldots$, $N$).  We adopt open boundary conditions with $n=1$ and $n=N$ being the two end sites. We mention that, although we explicitly discuss Fermion hopping in tight binding lattices as appropriate for band electrons in solids,  all our conclusions are equally valid for the corresponding bosonic case in atomic optical or photonic lattices since we are considering a noninteracting 1D quantum system. The first term in the Hamiltonian is the kinetic energy from the  nearest-neighbor hopping, and the second describes the on-site potential energy. We have derived this model starting from a special 3D lattice model with a tilted magnetic field. The details of this derivation are given in the Appendix.~\ref{appendix}. 

 All the parameters in the 1D model (Eq.~(\ref{aah1dm})) are independently tunable from an experimental perspective, irrespective of how they are obtained starting from the 3D model. In this sense, the 1D model is more flexible and general from an experimental point of view.  In the rest of this paper we will work with this effective 1D AAH Hamiltonian with two phase parameters. The hopping strength ($t$, $\lambda_{y\ (z)}$ $\lambda_{xy\ (xz)}$), phases ($\phi^{d}_{y\ (z)},\phi^{od}_{y (z)}$), and flux fractions ($b_y, b_z$) appearing in the Hamiltonian can be treated as controllable parameters in the experimental context. 
 
 A simpler version of the incommensurate AAH model has also been successfully implemented in photonic waveguide experiments\cite{lahini08,lahini2009}. The phase of the onsite modulation assumes the role of the propagation direction of  light injected in to the waveguide.  Photonic waveguides simulating incommensurate AAH model have realized the 1D localization-delocalization transition \cite{lahini08} and topological pumping of light\cite{kraus1,verbin13}. Topological aspects of an incommensurate AAH model have been discussed and debated in Refs.~\onlinecite{kraus1,kraus2,verbin13,brouwer,srirammm} and do not concern us at all in the current work. We emphasize that all our analysis and results are based on the physics of the commensurate $\pi$-flux AAH model and not the incommensurate 1D model. 

\section{Topological semimetal in 3D with inversion symmetry}\la{cssm}
  %%%%%%%%%%%%%%%%%%%%%%%%%%%%%%%%%%%%%%%%%%%%%%%%%%%%%%%%%%%%%%%%%%%%%%%%%%%%%%%%%%%%%%%%%%?
\begin{center}
\begin{figure}[htb!]
\textrm{(a)}\hspace{4cm}\textrm{(b)}
\includegraphics[width=4cm,height=2.5cm]{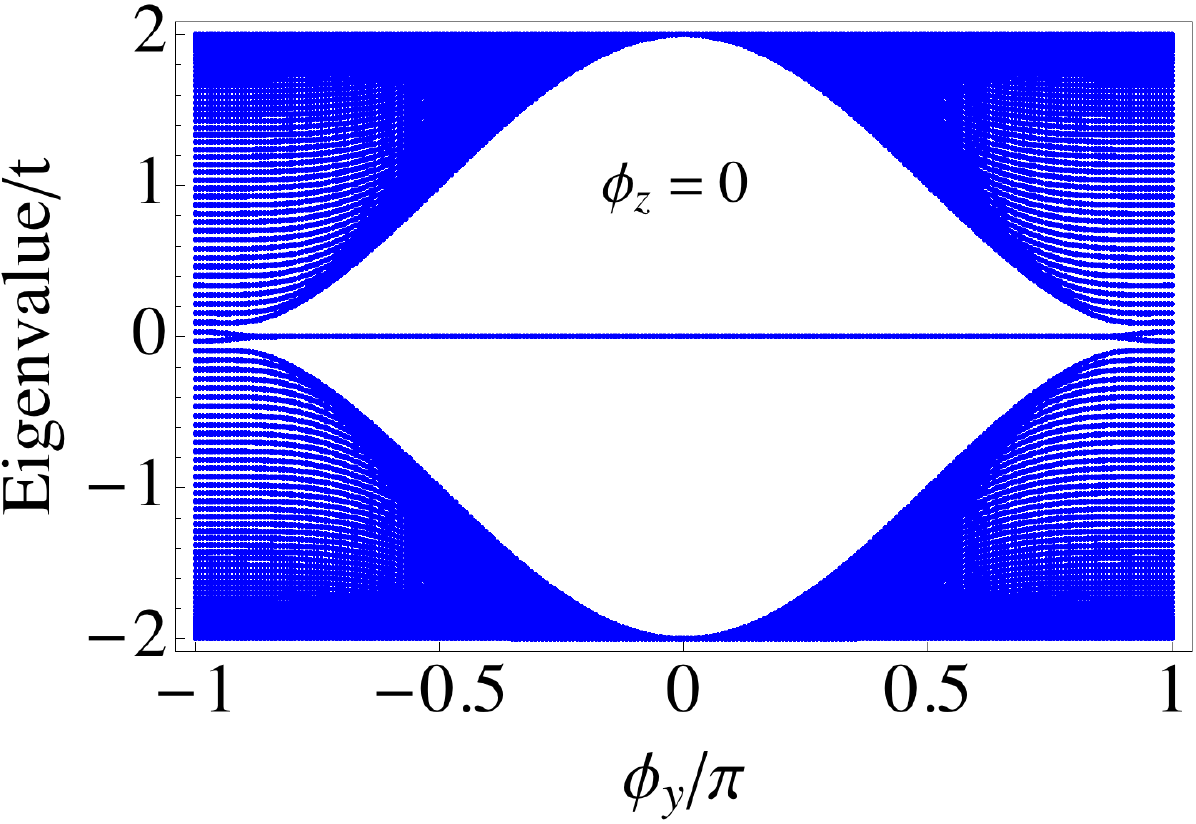}\hspace{0.1cm}\includegraphics[width=4.0cm,height=2.5cm]{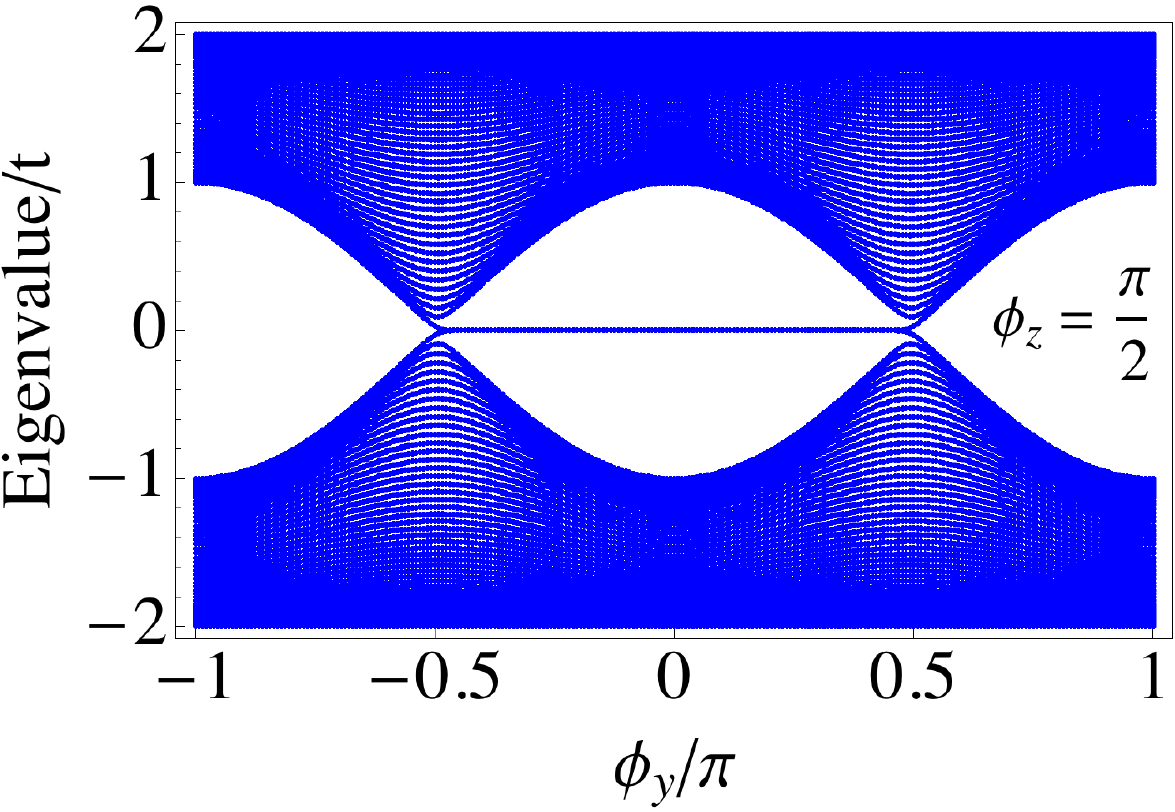}\\
\hspace{-0.2cm}\textrm{(c)}\hspace{4cm}\textrm{(d)}
\includegraphics[width=4.0cm,height=2.5cm]{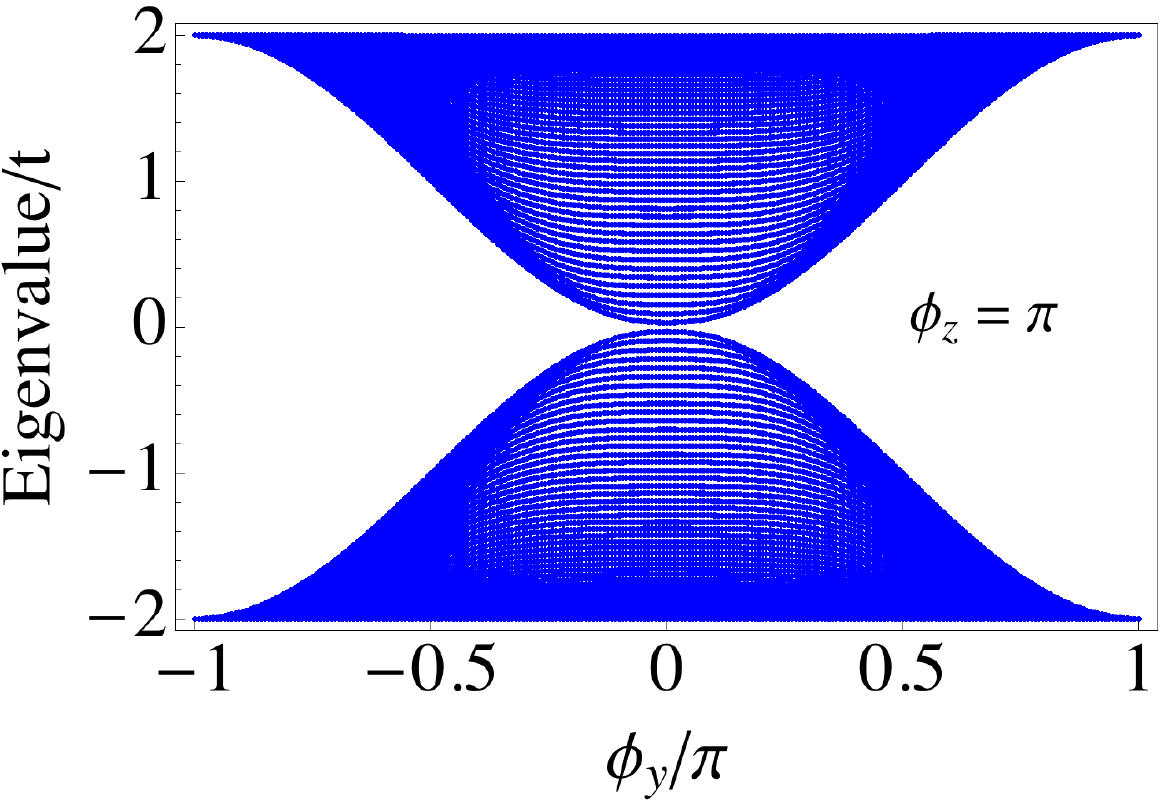}\hspace{0.1cm}\includegraphics[width=4.0cm,height=2.5cm]{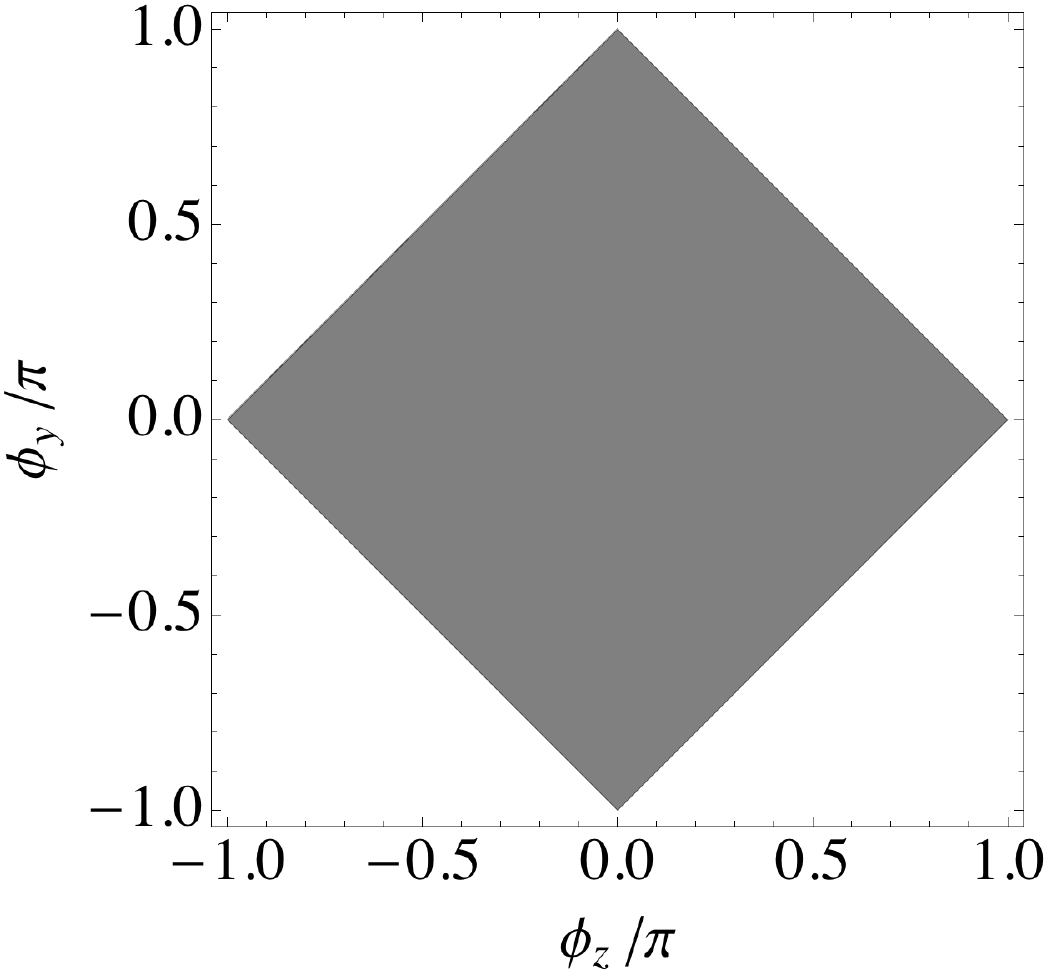}
\caption{ Numerical Energy spectrum for tight binding chain of 100 sites with open boundary conditions and parameter value $\lambda_{xy}=\lambda_{xz}=0.3$,  $\lambda_{y}=\lambda_{z}=0$, $t=1$. Plots  (a), (b) and (c) shows the evolution of the spectrum as a function of $\phi_y$ for $\phi_z=0, \pi/2, \pi$ respectively. (d)Region plot for the inequality condition $|\Delta_+(\phi_y,\phi_z)|>|\Delta_-(\phi_y,\phi_z)|$ as a function of  for existence of zero energy modes. The shaded region corresponds to the zero energy states and the boundary of the region corresponds to the topological phase transition where the zero energy states terminate as seen in (a), (b) and (c).}
\label{od}
\end{figure}
\end{center}
%%%%%%%%%%%%%%%%%%%%%%%%%%%%%%%%%%%%%%%%%%%%%%%%%%%%%%%%%%%%%%%%%%%%%%%%%%%%%%%%%%%%%%%%%
In this section, we start from the simplest case of the off-diagonal AAH model with two cosine modulations. This case corresponds to the parameter values of  $\lambda_{y\ (z)}=0$ and consequently has full inversion symmetry in 3D. 

The off diagonal AAH equation with two phase parameters can be written as,
\bea
H=\sum^{N-1}_{n=1}[t(1&+&\lambda_{xy}g_n(b_z,\phi_y)+\lambda_{xz}g_n(b_y,\phi_z)]c_{n+1}^{\dagger}c_{n}\nonumber\\
&+&H.c.]
\la{aahod}
 \eea
In the above equation we have set $\phi^{d}_{y\ (z)}=\phi^{od}_{y (z)}=\phi_{y (z)}$ without loss of generality.  The above Hamiltonian reduces to a 1D off-diagonal Hamiltonian for $b_z=b_y=1/2$, which was recently\cite{sriramaah,li} mapped to decoupled Majorana chains that manifest topological zero energy modes and support a $Z_2$ topological index~\cite{kitaev01}. $\lambda_{xz}\ne \lambda_{xy}\ne0$ provides generalized conditions for the existence of the zero energy modes. Following Ref.~(\onlinecite{sriramaah}), we rewrite the Hamiltonian in Eq.~(\ref{aahod}) in the Majorana basis for $b_z=b_y=1/2$. We define $c_{2n}=\gamma_{2n}+i\tau_{2n},\  c_{2n+1}=\tau_{2n+1}+i\gamma_{2n+1}$, where $\gamma$ and $\tau$ are two species of Majorana Fermions. In this new basis, Eq.~(\ref{aahod}) becomes,
\bea
H&=&\sum_{n}[\Delta_{-}(\phi_y,\phi_z)(\gamma_{2n}\gamma_{2n-1})+\Delta_{+}(\phi_y,\phi_z)(\gamma_{2n}\gamma_{2n+1})]\nonumber\\
&-&\sum_{n}[\Delta_{-}(\phi_y,\phi_z)(\tau_{2n}\tau_{2n-1})+\Delta_{+}(\phi_y,\phi_z)(\tau_{2n}\tau_{2n+1})],\nonumber\\
 \la{offdmajorana}
\eea
where $\Delta_{\pm}(\phi_y,\phi_z)=2it(1\pm\lambda_{xy}\cos\phi_{y}\pm\lambda_{xz}\cos\phi_{z})$. For $|\Delta_+(\phi_y,\phi_z)|>|\Delta_-(\phi_y,\phi_z)|$, the Majorana chain is topologically nontrivial and has one zero-energy Majorana mode localized at each end\cite{sriramaah}. For the opposite regime $|\Delta_+(\phi_y,\phi_z)|<|\Delta_-(\phi_y,\phi_z)|$, the system is topologically trivial with no end modes. In Fig.~(\ref{od}), we show the numerical energy spectrum as a function of one of the phase parameters $\phi_y$ and monitor the spectral evolution as a function of $\phi_z$. In Fig.~(\ref{od}a, b, c) we see that the zero energy modes exist between the band touching points. The band touching points can be moved using the second phase parameter $\phi_z$. Fig.~(\ref{od}c) shows maximal separation of band touching points for $\phi_z=0$ connected by degenerate zero energy modes. For $\phi_z=\pi$, Fig.~(\ref{od}c)  shows that the two band touching points merge thereby collapsing the line of zero energy modes to a single band touching point. We plot the regions in $(\phi_y,\phi_z)$ plane which satisfy the condition $|\Delta_+(\phi_y,\phi_z)|>|\Delta_-(\phi_y,\phi_z)|$ for the existence of the zero energy modes (see Fig.~(\ref{od}) d). The zero mode existence condition in the $(\phi_y,\phi_z)$ plane can be interpreted as the zero energy Fermi surface of the 3D band structure. 

\subsection{Discrete symmetries and robustness}
The topology of the 1D model defined in Eq.~(\ref{aahod}) is associated with the chiral symmetry. Chiral symmetry arises as a combination of the time reversal and particle hole symmetry. Under time-reversal transformation, a $\pi$ flux simply turns into a $-\pi$ flux. Since the magnetic flux terms are only well defined modulo $2\pi$ for a lattice, the system is then invariant under the time-reversal transformation. Therefore, the system shows no QH effect and thus has no QH edge modes. In fact, this lattice model has no band gap but contains two Dirac points with linear dispersion in analogy to graphene.  The invariance of the 1D model under $c_n\rightarrow (-1)^n c^\dagger_n$ and $c_n^\dagger \rightarrow (-1)^n c_n$ demonstrates particle hole symmetry. In terms of the Altland-Zirnbauer classification the 1D topology belongs to the BDI class\cite{ryu09} which has an integer ($Z$) invariant. Since we only allow for short range (nearest-neighbor) hopping terms in our model, the integer Z index can only take values 0 or 1 (i.e., $Z_2$). If the kinetic energy is dominated by the longer-range hopping terms, a higher topological index can be achieved. We do not consider longer range hopping in this paper.  As long as the particle-hole symmetry is preserved~\cite{ryu2002}, the mapping to two decoupled Majorana chains remains valid and thus the end modes are stable with their energy pinned to zero. The robustness of the zero energy end modes on weakly breaking the particle hole symmetry has been established in Ref.~(\onlinecite{sriramaah}). The onsite modulations breaking the inversion symmetry (time reversal) has a physical role in 3D. We will be considering such terms in detail in Sec.~\ref{weyl}.

  \section{Topological semimetal in 3D with broken inversion symmetry}\la{weyl}
 In this section, we break the inversion symmetry by adding onsite modulation terms.  Onsite cosine terms with $\pi$-flux break the inversion symmetry (while retaining the particle-hole symmetry).  We can write down general conditions for the existence of 3D TS phase for the general model in Eq.~(\ref{aah3d2a}) in the $\pi$-flux state $(b_z=b_y=\frac{1}{2})$. The general 1D Hamiltonian in the $\pi$-flux state can be written as,
  \bea
 H&	=&\sum^{N-1}_{n=1}(t(1+(-1)^n\lambda_{xy}\cos\phi_y+(-1)^n\lambda_{xz}\cos\phi_z))c_{n+1}^{\dagger}c_{n}\nonumber\\
		&+&H.c+\sum^{N}_{n=1}(-1)^n(\lambda_{y}\cos\phi_y+\lambda_{z}\cos\phi_z)) c_{n}^{\dagger}c_{n}.
		\la{aah3d2a}
\end{eqnarray}
To outline the symmetries of this 1D Hamiltonian, we can write it in the Rice-Mele\cite{ricemele} form,
\bea
 H&	=&\sum^{N-1}_{n=1}\Delta_{+}(\phi_y,\phi_z)a_{n}^{\dagger}b_{n}+\Delta_{-}(\phi_y,\phi_z)a_{n}^{\dagger}b_{n-1}+H.c.\nonumber\\
		&+&\sum^{N}_{n=1}\epsilon(\phi_y,\phi_z) (a_{n}^{\dagger}a_{n}- b_{n}^{\dagger}b_{n}).
		\la{ricemele}
		\eea
Where we have defined $\Delta_{\pm}(\phi_y,\phi_z)= t(1\pm\lambda_{xy}\cos\phi_{y}\pm\lambda_{xz}\cos\phi_{z})$ and $\epsilon(\phi_y,\phi_z)=(\lambda_{y}\cos\phi_y+\lambda_{z}\cos\phi_z)$.		
The corresponding Bloch Hamiltonian $h$ in the bipartite sub-lattice basis\cite{bloch}  can be written as,
\begin{gather}
h(k,\phi_y,\phi_z) = \bpm \epsilon(\phi_y,\phi_z) & \Delta(k,\phi_y,\phi_z) \\ \Delta^*(k,\phi_y,\phi_z)&-\epsilon(\phi_y,\phi_z) \epm,
\la{bloch}
\end{gather}		
where $\Delta(\phi_y,\phi_z)=\Delta_{+}e^{ikd/2}+\Delta_{-}e^{-ikd/2}$ and $d$ is the lattice constant. Note that $\epsilon(\phi_y,\phi_z)=0$ for some values of $(\phi_y,\phi_z)$ corresponds to an inversion symmetric (where the center of inversion is defined about a bond connecting two sites~(see Fig.~(\ref{exp}))) 1D chain and the topological invariant is given in terms of the quantized Zak phase (0 or $\pi$)\cite{bloch}. The BZ space satisfying the constraint $\epsilon(\phi_y,\phi_z)=0$ can be deconstructed into topological 1D slices with Zak phase either 0 or $\pi$.  Based on this principle we can develop a general algorithm for identifying topologically protected band touching points and surface states of a 3D TS or WSM,
\begin{itemize}
\item[I.] Identify points $(\phi_y,\phi_z)$ for which onsite terms (diagonal term $\epsilon(\phi_y,\phi_z)=0$  in Eq.~(\ref{bloch})) breaking the inversion symmetry vanish. The set of inversion symmetric points  $(\phi_y,\phi_z)$  admits exact mapping to double Majorana chains.
\item[II.]  Within the subspace of inversion symmetric points ($\epsilon(\phi_y,\phi_z)=0$), the conditions for the existence of topological zero energy modes can be identified from the zero mode existence condition of double Majorana chain as shown in Eq.~(\ref{offdmajorana}) (or Zak phase of $\pi$).
\end{itemize}
The above criterion is closely related to the Shockley criterion for the end states of a 1D chain and our model can be thought of as a generalization of the Shockley criterion for surface states of topological insulators to the case of WSM~\cite{shockley}. 
 %%%%%%%%%%%%%%%%%%%%%%%%%%%%%%%%%%%%%%%%%%%%%%%%%%%%%%%%%%%%%%%%%%%%%%%%%%%%%%%%%%%%%%%%%%?
\begin{center}
\begin{figure}[htb!]
\textrm{(a)}\hspace{4cm}\textrm{(b)}
\includegraphics[width=4cm,height=2.5cm]{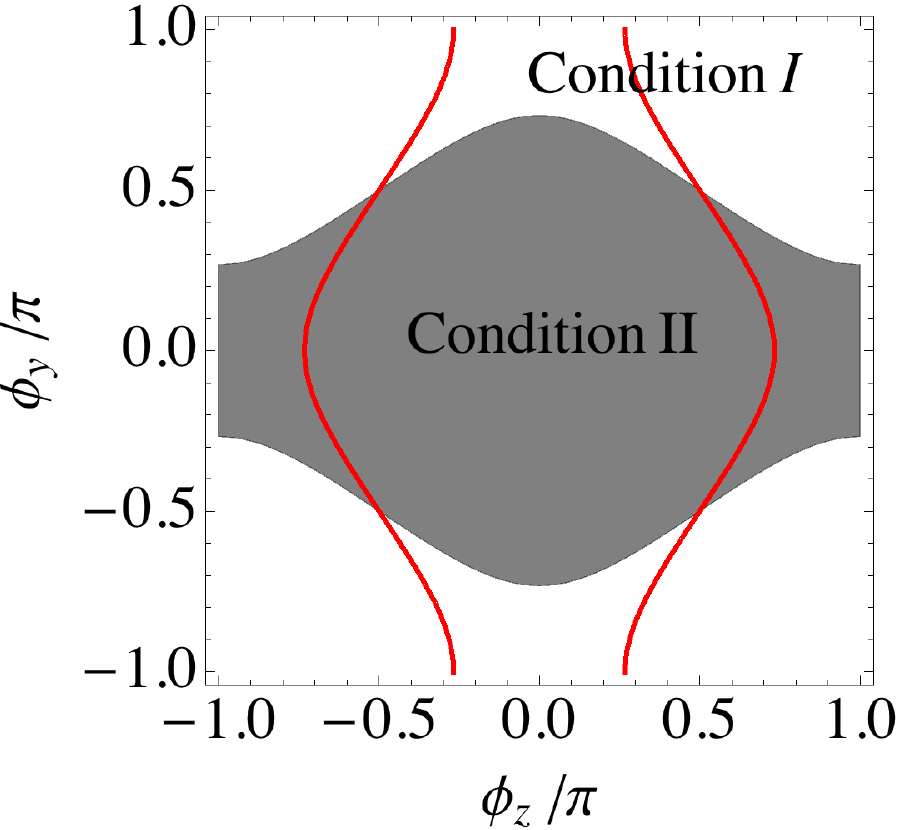}\hspace{0.1cm}\includegraphics[width=4.0cm,height=2.5cm]{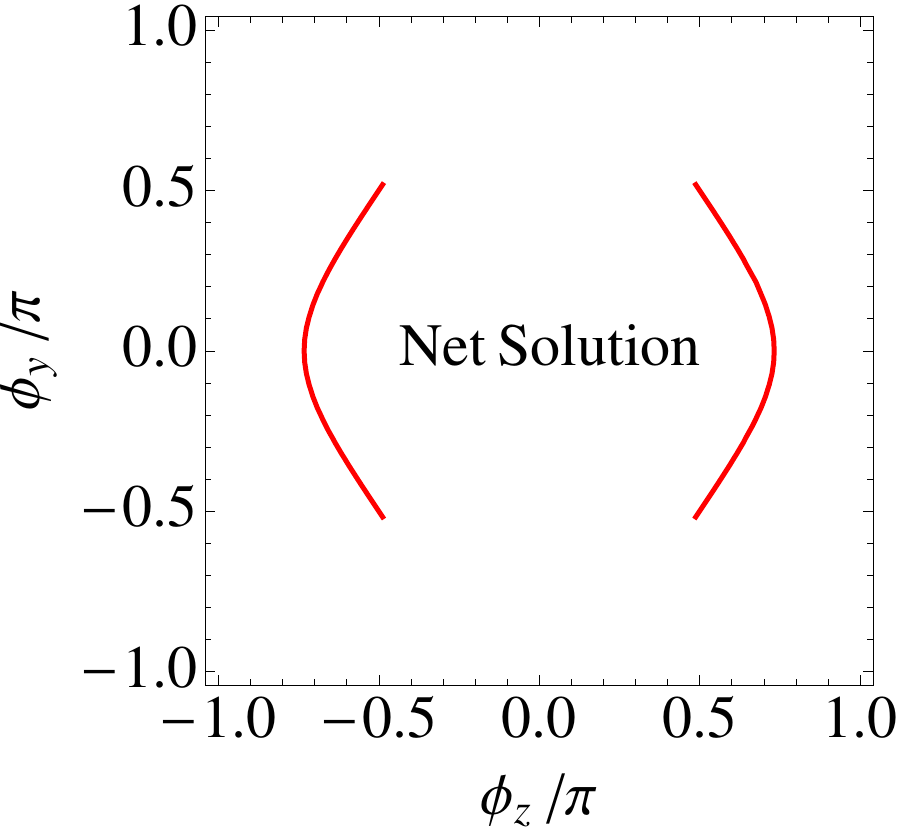}
\caption{Solution of $(\phi_y,\phi_z)$ satisfying Eqs.~(\ref{mastercond1} and \ref{mastercond2}) for $\lambda_y=\lambda_{xz}=0.6$, $\lambda_z=\lambda_{xy}=0.4$ and $t=1$.  (a) The red line denotes the contour satisfying the first equality condition of Eq.~(\ref{mastercond1}) for which the onsite modulation vanishes. Shaded region denotes the Brillouin zone patch satisfying the second inequality condition in Eq.~(\ref{mastercond2}) for the existence of the zero modes. (b) The red line corresponds to the zero energy (Weyl) arcs which is the region satisfying the intersection of both constraints. }
\label{master}
\end{figure}
\end{center}
%%%%%%%%%%%%%%%%%%%%%%%%%%%%%%%%%%%%%%%%%%%%%%%%%%%%%%%%%%%%%%%%%%%%%%%%%%%%%%%%%%%%%%%%%
The above set of criteria translates to the following conditions for the existence of topologically protected zero energy modes (band touching points),
\bea
\lambda_{y}g_n(b_z,\phi_y)+\lambda_{z}g_n(b_y,\phi_z)=0\\
|t(1+\lambda_{xy}g_n(b_z,\phi_y)+\lambda_{xz}g_n(b_y,\phi_z))|> \nonumber\\
|t(1-\lambda_{xy}g_n(b_z,\phi_y)-\lambda_{xz}g_n(b_y,\phi_z))|
\eea
These conditions are not limited to the $\pi$-flux case and can be generalized for any flux fraction $ b_y=b_z=1/(2q)$. The zero mode existence condition generalizes for b=1/(2q) as shown in Ref.~(\onlinecite{sriramaah}). 
 Writing out the above conditions explicitly for  $ b_y=b_z=1/2$ we obtain,
 \bea
&&\lambda_{y}\cos(\phi_y)+\lambda_{z}\cos(\phi_z)=0\la{mastercond1}\\
&&|1+\lambda_{xy}\cos(\phi_y)+\lambda_{xz}\cos(\phi_z)|>\nonumber\\
&&|1-\lambda_{xy}\cos(\phi_y)-\lambda_{xz}\cos(\phi_z)|
\la{mastercond2}
\eea
In Fig.~(\ref{master}), we present an example satisfying the above conditions for the parameter values  $\lambda_y=\lambda_{xz}=0.6$, $\lambda_z=\lambda_{xy}=0.4$ and $t=1$. In Fig.~(\ref{master}a) the red line shows the locus of $(\phi_y, \phi_z)$ satisfying Eq.~(\ref{mastercond1}) corresponding to the vanishing of the onsite modulations. Note that in three dimensions, these points are topologically unavoidable. The shaded region corresponds to the region of $(\phi_y, \phi_z)$ for which the double Majorana chain admits zero energy solutions. Fig.~(\ref{master}b) plots the intersection of these conditions that constraints the topological band touching points to arcs in the zero energy Fermi surface thus giving rise to the Weyl arcs for the 3D WSM.  

 \subsection{Example I: $\lambda_{z}\ne0,\ \lambda_{xy}\ne0\  \lambda_{y}=\lambda_{xz}=0$}\la{2dcs}
  In this part, we forbid NN hopping in the $y$ direction and NNN hopping in the $xz$ plane as a starting example, then allow them to have non-zero values in the next sections.  From the point of view of the cubic lattice, this case is equivalent to stacking inversion symmetric 2D planes and gluing them via NN hopping in the z direction along which the inversion symmetry is broken. 
 \begin{eqnarray}
H&=&\sum^{N-1}_{n=1}(t(1+(-1)^n\lambda_{xy}\cos(\phi_y))c_{n+1}^{\dagger}c_{n}+h.c.\nonumber\\
		&+&	\sum^{N}_{n=1}\lambda^{z}(-1)^n\cos(\phi_z) c_{n}^{\dagger}c_{n}.
		\la{aah3d}
		  \end{eqnarray}
		   %%%%%%%%%%%%%%%%%%%%%%%%%%%%%%%%%%%%%%%%%%%%%%%%%%%%%%%%%%%%%%%%%%%%%%%%%%%%%%%%%%%%%%%%%%?
\begin{center}
\begin{figure}[htb!]
\textrm{(a)}\hspace{4cm}\textrm{(b)}
\includegraphics[width=4cm,height=2.5cm]{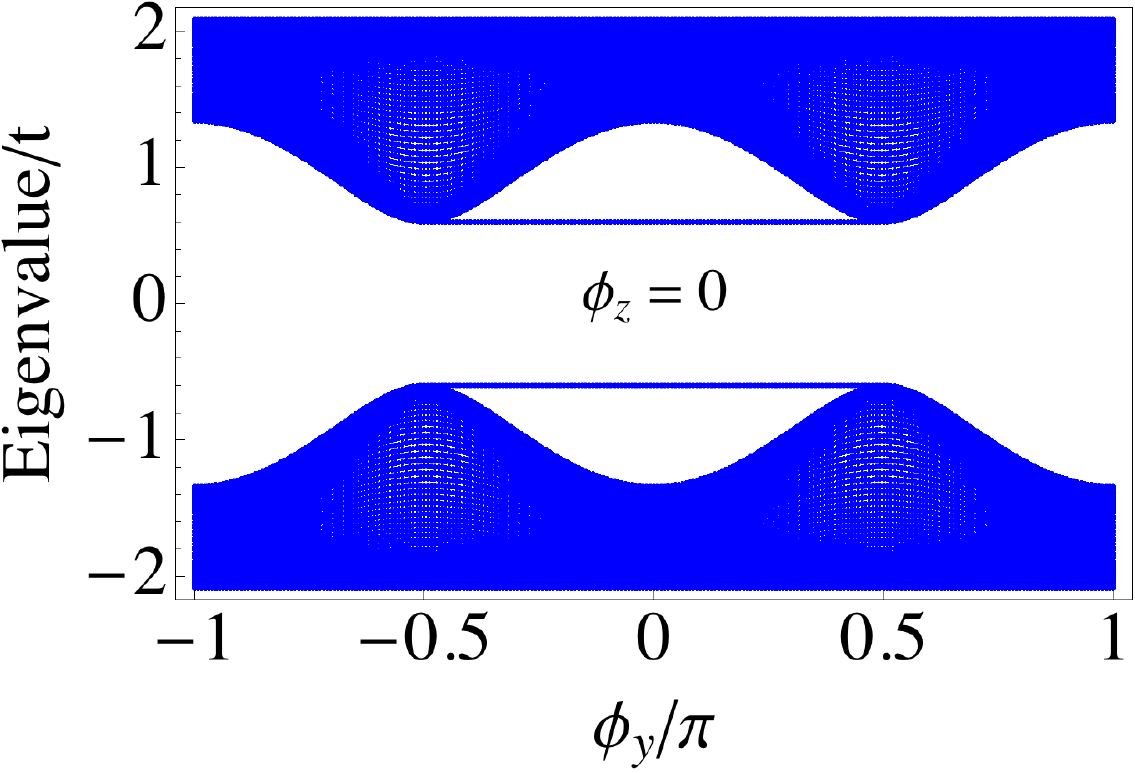}\hspace{0.1cm}\includegraphics[width=4.0cm,height=2.5cm]{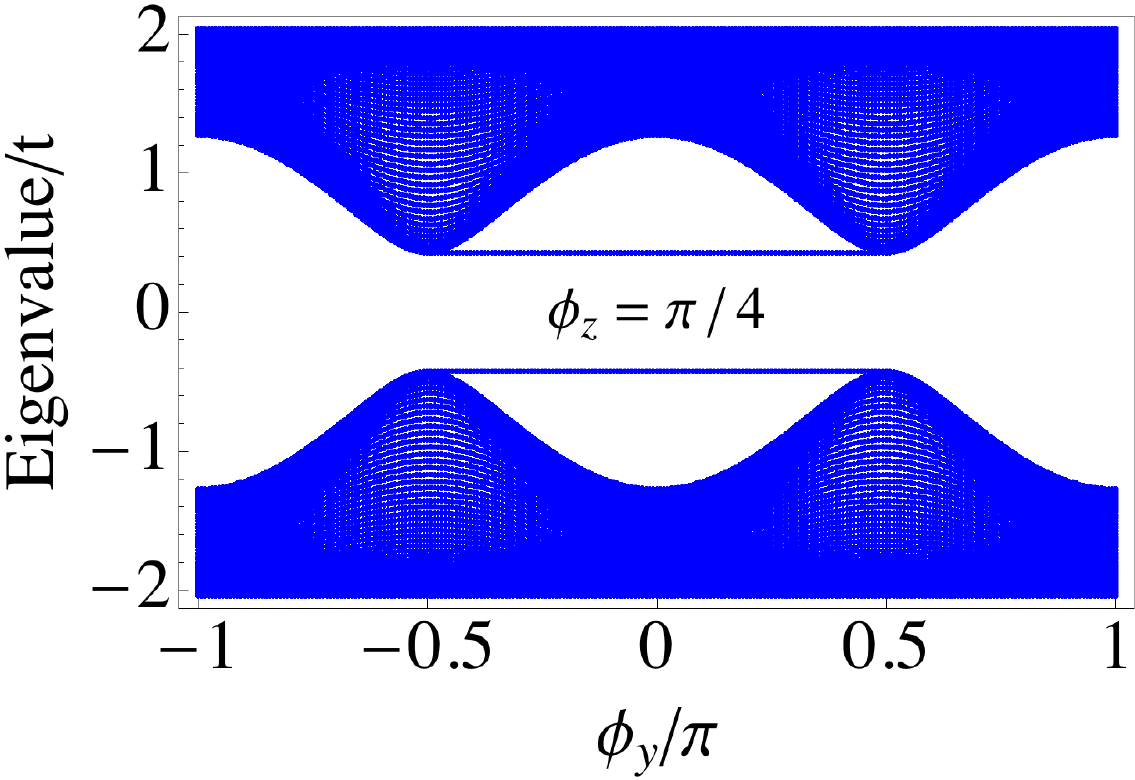}\\
\hspace{-0.2cm}\textrm{(c)}\hspace{4cm}\textrm{(d)}
\includegraphics[width=4.0cm,height=2.5cm]{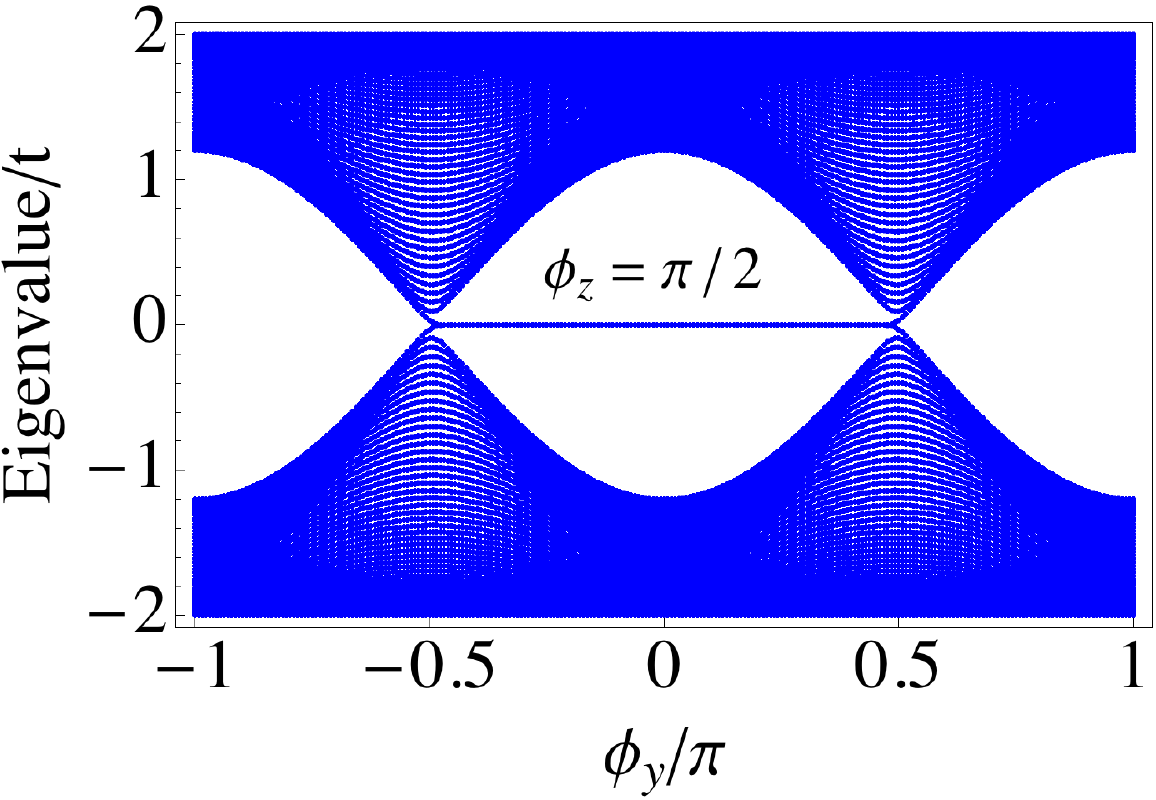}\hspace{0.1cm}\includegraphics[width=4.0cm,height=2.5cm]{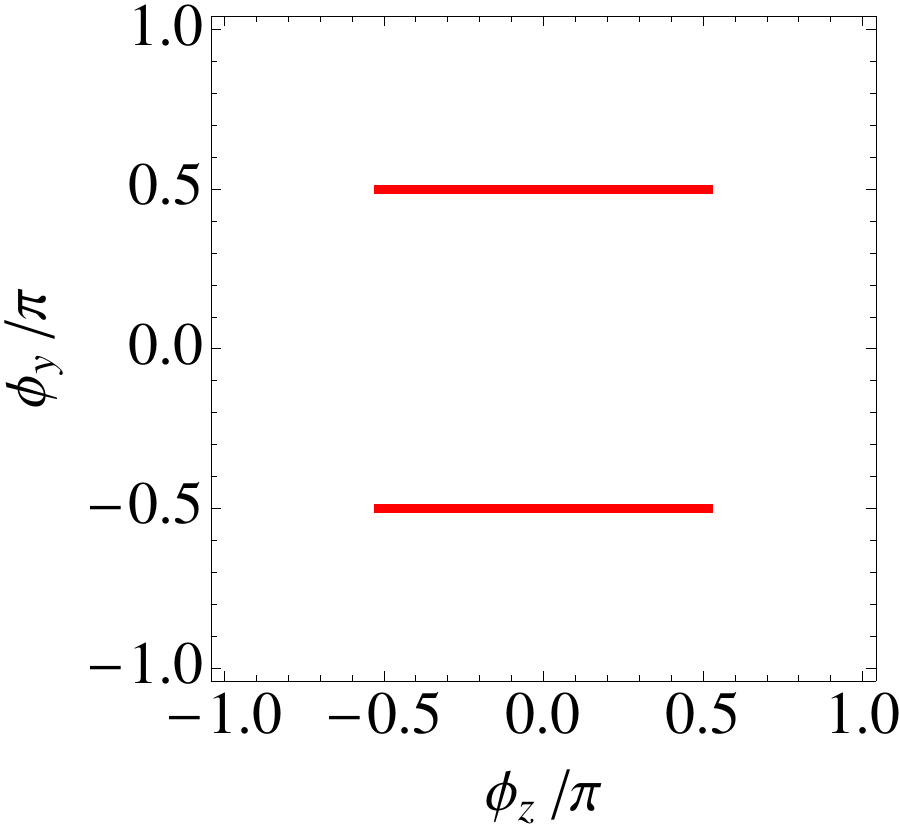}
\caption{ Numerical Energy spectrum for tight binding chain of 100 sites with open boundary conditions and parameter value $\lambda_{z}=\lambda_{xy}=0.3$,  $\lambda_{y}=\lambda_{xz}=0$, $t=1$. Plots  (a), (b) and (c) show the evolution of the spectrum as a function of $\phi_y$ for $\phi_z=0, \pi/4, \pi/2$ respectively. (d)Region plot for the inequality condition $|\Delta_+|>|\Delta_-|$ as a function of $(\phi_y,\phi_z)$ for existence of zero energy modes. The red lines corresponds to the zero energy states and the ends of the lines corresponds to the topological phase transition point where the zero energy states terminate as seen in (a), (b) and (c).}
\label{2dcs}
\end{figure}
\end{center}
%%%%%%%%%%%%%%%%%%%%%%%%%%%%%%%%%%%%%%%%%%%%%%%%%%%%%%%%%%%%%%%%%%%%%%%%%%%%%%%%%%%%%%%%%
This simple form of the Hamiltonian enables us to directly deconstruct the topological features in the 3D Brillouin zone. Each value of $(\phi_y,\phi_z)$ corresponds to a {\it specific} 1D Hamiltonian. Note that the coordinates of $(\phi_y,\phi_z)$ at which the zero energy modes occur correspond to the set of 1D chains that are topologically non-trivial with an invariant that is associated with the two decoupled Majorana chains of Eq.~(\ref{offdmajorana}). This is due to the fact that the 3D TS acquires its topological features from a lower dimensional (1D) symmetry protecting the topology.  The interesting feature of the 3D physics here is that the topologically non-trivial slices of BZ cannot be removed by breaking the symmetries that protect the invariant. The addition of the onsite term coming from the z direction hopping of the cubic lattice breaks the inversion symmetry along this direction and opens a gap in the spectrum. This spectrum is gapped everywhere except for the points in the Brillouin zone that satisfy the conditions (Eqs.~\ref{mastercond1} and \ref{mastercond2})
\be
\cos(\phi_z)=0,\ \ \cos(\phi_y)\ge0,
\la{cond1}
\ee
for which the BZ slice is inversion symmetric.
For the points $\phi_z=\pm\frac {\pi}{2}$, $H$ is inversion symmetric and admits mapping to the Majorana basis as shown in Eq.~(\ref{offdmajorana}). The conditions for the existence of Majorana zero modes are given in Eq.~(\ref{cond1}). Thus the set of constraints $\phi_z=\pm\frac {\pi}{2}$ and $\frac {\pi}{2}<\phi_y<\frac {\pi}{2}$ define the zero energy Fermi surface in the 3D BZ. One can immediately recognize that the Fermi surface is reduced to zero energy lines (nodal lines) that connect the band touching points similar to that of nodal semimetal. In Fig.~(\ref{2dcs}), we plot the numerical energy spectrum of the 1D AAH model given in Eq.~(\ref{aah3d}) for a tight binding chain of 100 sites with open boundary conditions and the parameter values $\lambda_{z}=\lambda_{xy}=0.3$,  $\lambda_{y}=\lambda_{xz}=0$, $t=1$. The topologically trivial (gapped) 2D layers ($\phi_z\ne\pm\pi/2$) manifest dispersionless  states or flat bands (see Fig.~(\ref{2dcs} a and c)). This case been identified as the `topological nodal semimetals' with line nodes in Ref.~(\onlinecite{balents11}) at which the valence band and the conduction band touch (see Fig.~(\ref{2dcs}b)). The nodal lines of zero modes are attributed to the underlying inversion symmetry in the 2D layer of the $(n,\phi_y)$ for the values of $\phi_z=\pm\pi/2$.
 \subsection{Example II: $\lambda_{z}\ne0,\ \lambda_{xy}\ne0\  \lambda_{y}\ne 0,\ \lambda_{xz}\ne0$}
 %%%%%%%%%%%%%%%%%%%%%%%%%%%%%%%%%%%%%%%%%%%%%%%%%%%%%%%%%%%%%%%%%%%%%%%%%%%%%%%%%%%%%%%%%%?
\begin{center}
\begin{figure}[htb!]
\textrm{(a)}\hspace{4cm}\textrm{(b)}
\includegraphics[width=4cm,height=2.5cm]{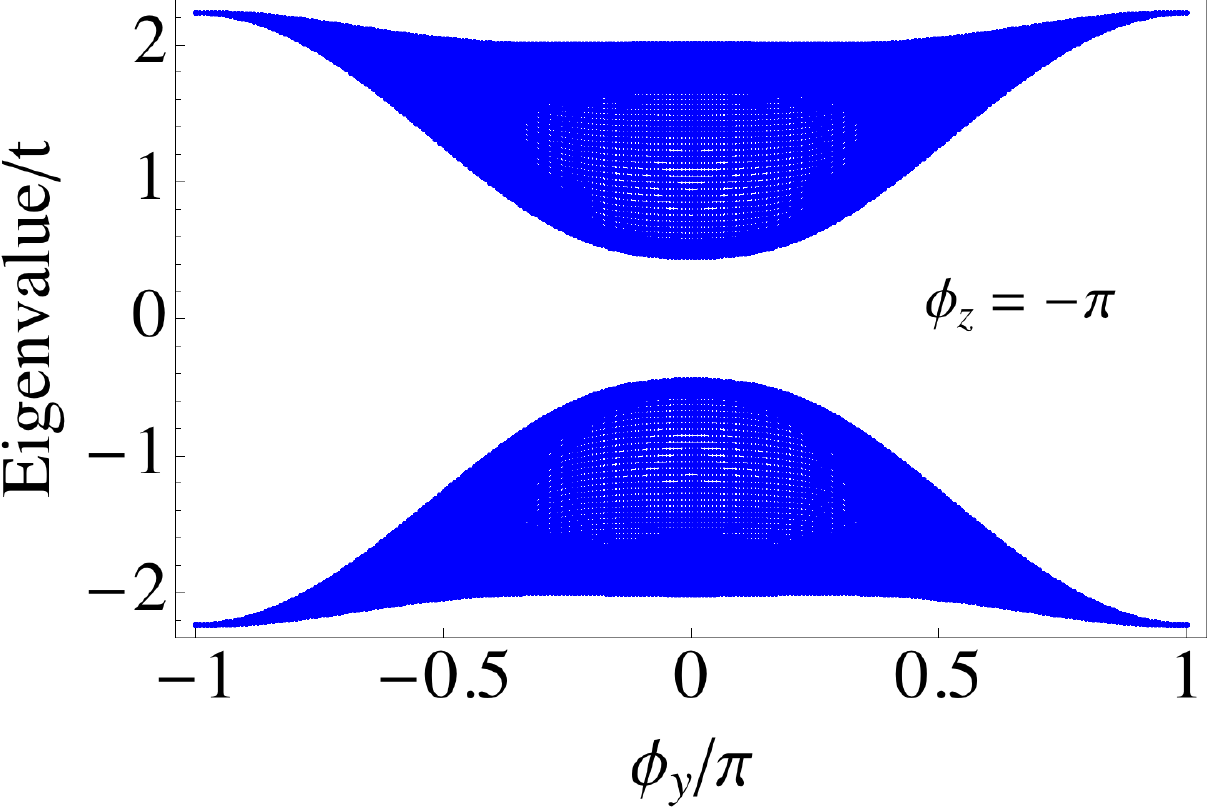}\hspace{0.1cm}\includegraphics[width=4.0cm,height=2.5cm]{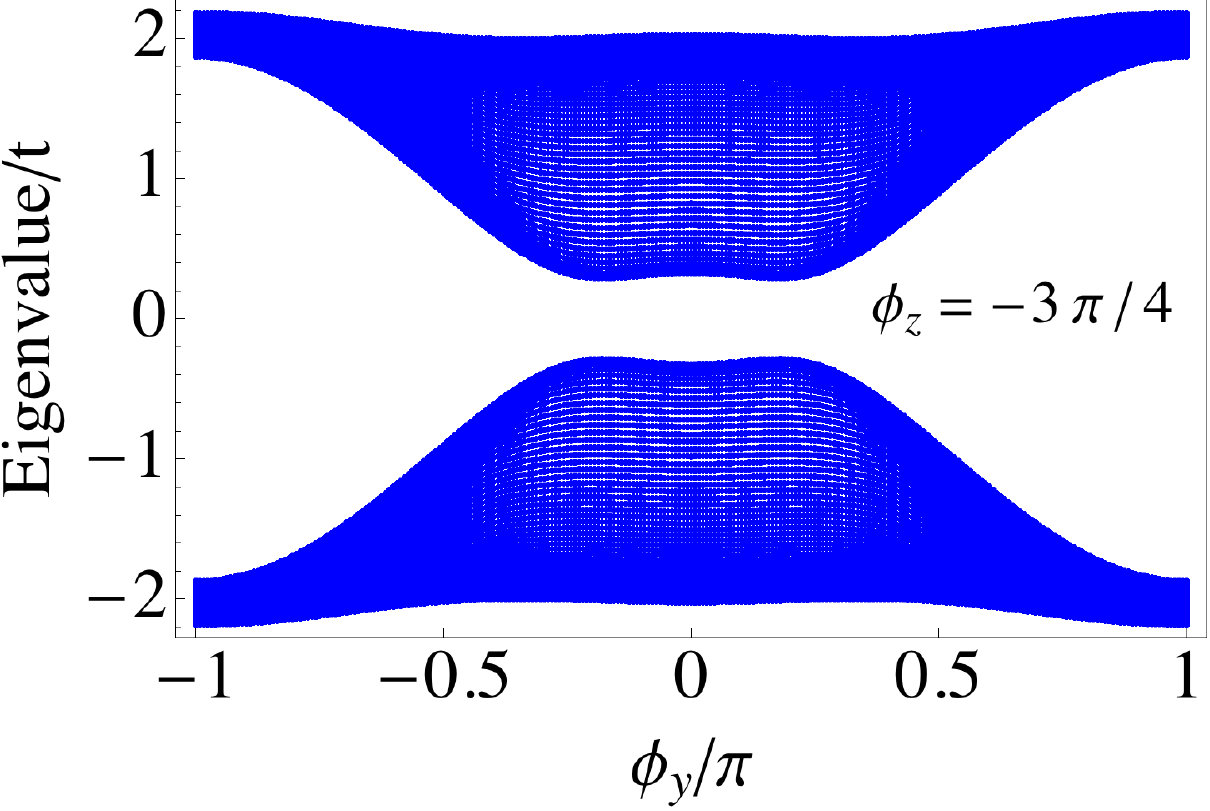}\\
\hspace{-0.2cm}\textrm{(c)}\hspace{4cm}\textrm{(d)}
\includegraphics[width=4.0cm,height=2.5cm]{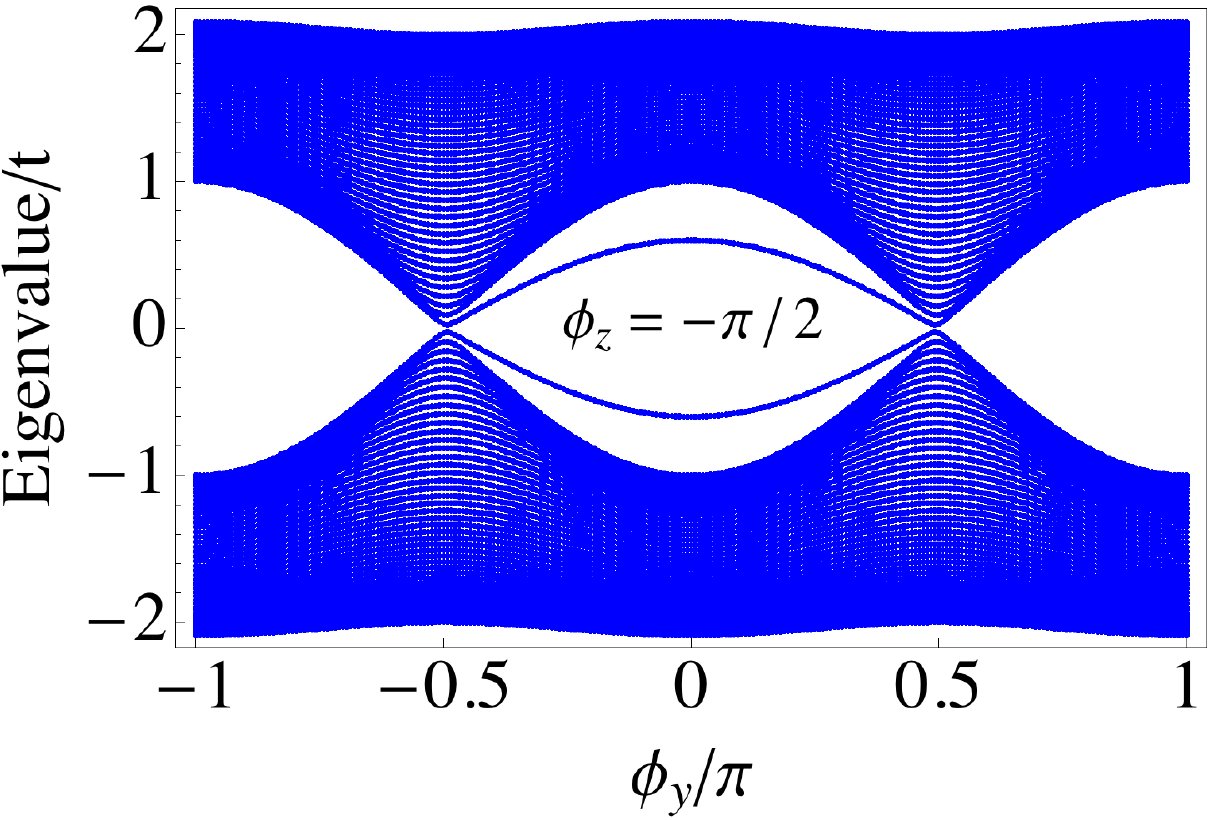}\hspace{0.1cm}\includegraphics[width=4.0cm,height=2.5cm]{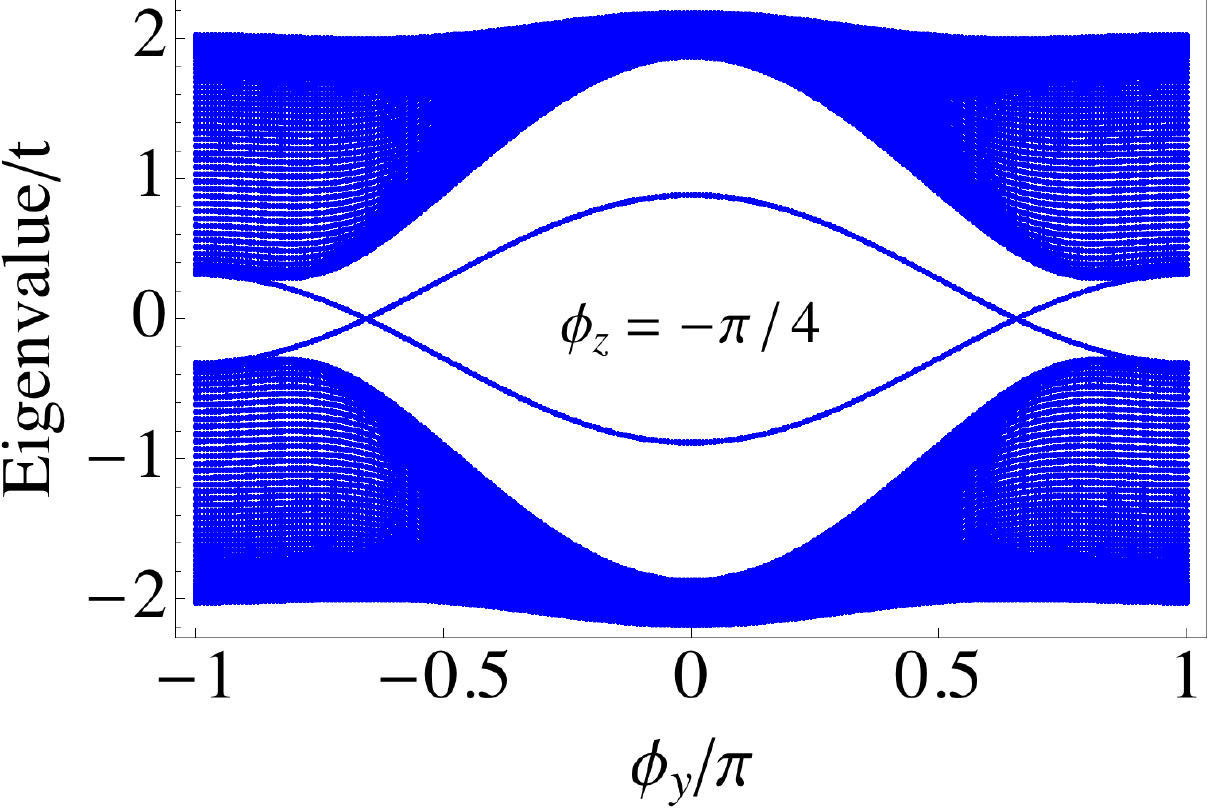}\\
\hspace{-0.2cm}\textrm{(e)}\hspace{4cm}\textrm{(f)}
\includegraphics[width=4.0cm,height=2.5cm]{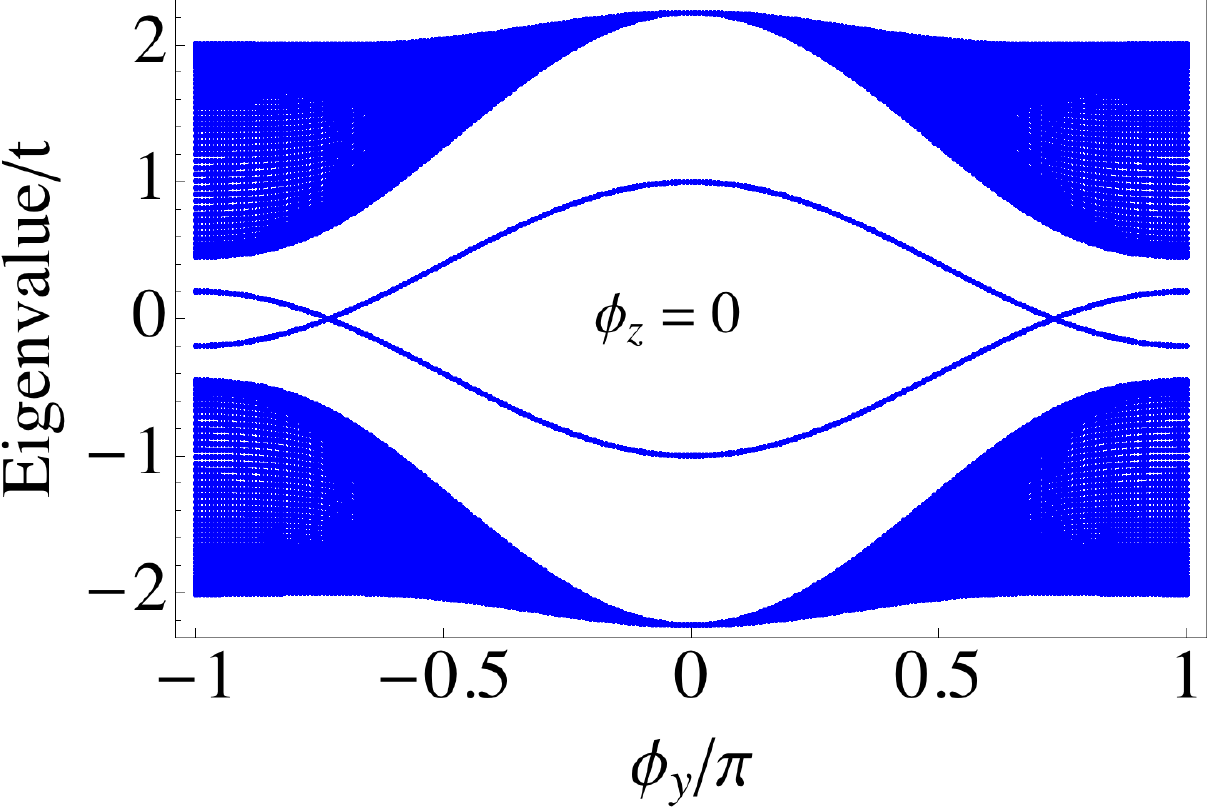}\hspace{0.1cm}\includegraphics[width=4.0cm,height=2.5cm]{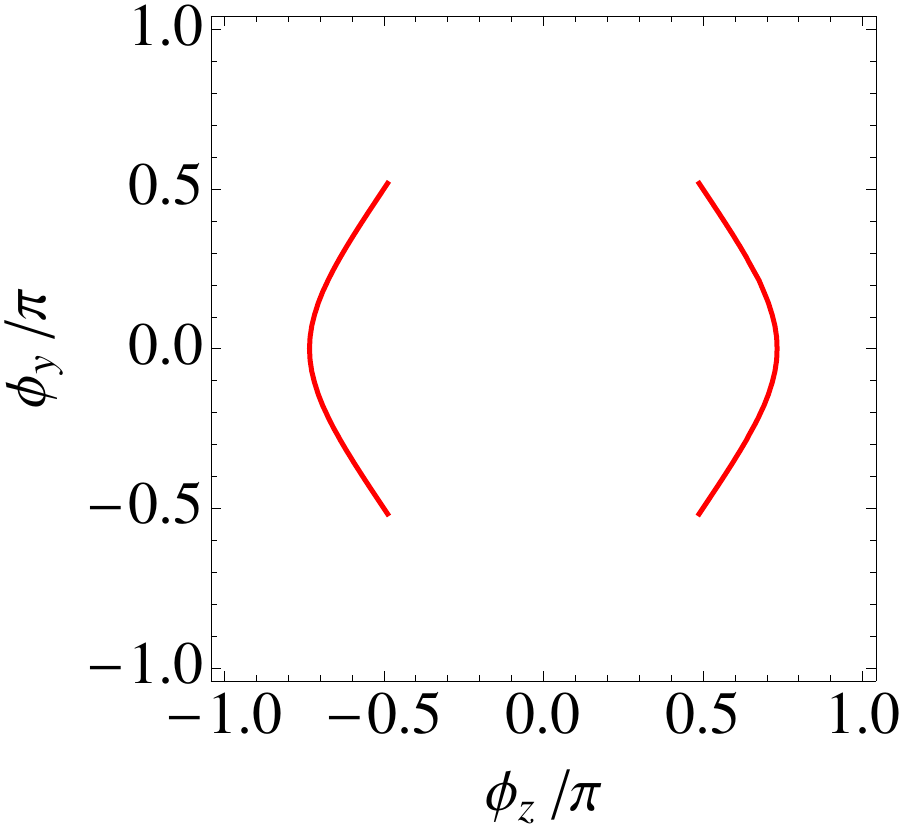}
%\hspace{-0.2cm}\textrm{(g)}\hspace{4cm}\textrm{(h)}
%\includegraphics[width=4.0cm,height=2.5cm]{fig4g.pdf}\hspace{0.1cm}\includegraphics[width=4.0cm,height=2.5cm]{fig4h.pdf}\\
%\hspace{-0.2cm}\textrm{(i)}\\
%\includegraphics[width=4.0cm,height=2.5cm]{fig4i.pdf}
\caption{ Numerical Energy spectrum for tight binding chain of 100 sites with open boundary conditions and parameter value $\lambda_{z}=\lambda_{xy}=0.4$,  $\lambda_{xz}=0.6,\ \lambda_{y}=0.4$, $t=1$. Plots  (a)-(b) show the evolution of the spectrum as a function of $\phi_y$ for $\phi_z=-\pi, -3\pi/4$ respectively.  Plots  (c)-(e) show the evolution of the spectrum as a function of $\phi_y$ for $\phi_z=-\pi/2, -\pi/4,\ 0$ respectively.  (f) Region plot tracing the locus of zero energy Fermi surface or the Weyl arc states for the condition given in Eq.~(\ref{mastercond1} and \ref{mastercond2}) as a function of $(\phi_y,\phi_z)$.  The red lines corresponds to the zero energy states and end of the line corresponds to the topological phase transition point where the zero energy states terminate or the Weyl nodes as seen in the spectrum plot (c).}
\label{chernsm}
\end{figure}
\end{center}
%%%%%%%%%%%%%%%%%%%%%%%%%%%%%%%%%%%%%%%%%%%%%%%%%%%%%%%%%%%%%%%%%%%%%%%%%%%%%%%%%%%%%%%%% 
Now we consider the general 3D AAH model for $b_y=b_z=\frac{1}{2}$ containing all terms appearing in Eq.~(\ref{aah1dm}). For this general case the Hamiltonian becomes 
 \bea
 H&	=&\sum^{N-1}_{n=1}(t(1+(-1)^n\lambda_{xy}\cos\phi_y+(-1)^n\lambda_{xz}\cos\phi_z))c_{n+1}^{\dagger}c_{n}\nonumber\\
		&+&H.c+\sum^{N}_{n=1}(-1)^n(\lambda_{y}\cos\phi_y+\lambda_{z}\cos\phi_z)) c_{n}^{\dagger}c_{n}.
		\la{aah3d2}
\end{eqnarray}
Using  methods developed in the previous section, we try to identify special points $(\phi_y,\phi_z)$ in the 3D Brillouin zone for which the 1D Hamiltonian defined by Eq.~(\ref{aah3d}) admits mapping to the double Majorana chain. In spite of breaking the inversion symmetry in the $xy$ plane there are ``topologically unavoidable" points where the zero energy states must appear. Solving the constraints in Eqs.~(\ref{mastercond1} and \ref{mastercond2}) for ($\phi_y,\phi_z$),  we obtain conditions for the existence of topologically protected zero energy modes. In addition to the one dimensional topological features, this case manifests a 2D topological transition in the three dimensional Brillouin zone. To understand this effect we plot the numerical energy spectrum in Fig.~(\ref{chernsm}) as a function of  one of the phases $\phi_y$ and monitor its evolution as a function of the other phase angle $\phi_z$. We fix $\lambda_y=\lambda_{xz}=0.4$ and $\lambda_{xy}=\lambda_z=0.6$ which indicates that the inversion symmetry is strongly broken in all planes. As a consequence we see a topological phase transition driven by $\phi_z$ between two distinct 2D phases. In Figs.~(\ref{chernsm}(a), (b)), we see a completely gapped energy spectrum corresponding to a normal spectrum for the 2D planes labelled by $\phi_z=-\pi,\ -3\pi/4, -\pi/2$. Figs.~(\ref{chernsm}(c-e)) describing a topologically non-trivial two mirror copies of a Chern insulator state (due to time reversal symmetry).  The mid-gap states of these insulator planes cross at zero energy and the topological nature of these degenerate points arises from the topological invariants associated with the double Majorana chain. This phase coincides with the existence of the topological zero energy modes and can be extracted by solving the conditions outlined in Eq.~(\ref{mastercond1} and \ref{mastercond2}). Topological insulator planes with time reversal symmetry are separated from the normal insulating planes by a  critical 2D plane where the pair of Weyl nodes appear in the 3D BZ shown in Fig.~(\ref{chernsm}c). Thus the 3D BZ separates in to trivial and non-trivial topological stacks of 2D insulating planes. For $-\pi<\phi_z<-\pi/2$ and $\pi/2<\phi_z<\pi$, the 2D planes corresponds to topologically trivial insulators. For $-\pi/2<\phi_z<\pi/2$, the 2D planes correspond to topologically non-trivial topological insulators which are same as two mirror copies of Chern insulators due to the presence of time reversal symmetry. $\phi_z=\pm\pi/2$ correspond to the critical semimetallic 2D plane.  Similar momentum space crossover between normal and topological insulators has been identified as `Chern semimetals' and has been predicted from first principles calculations\cite{chernsemimetal} in a ferromagnetic  compound $HgCr_2Se_4$. The topological planes for this case manifest Chern number$=\pm2$ which is different from our theory where the topological planes have Chern number $\pm1$ corresponding to each mirror copy of Chern insulator in the half Brillouin zone. In our theory, this Chern semimetal arises naturally as a consequence of constructing the 3D topological phase from generalized 1D 2-parameter AAH models and connecting the physics to the double Majorana chain systems with zero energy topological states. In the next section, we obtain topological invariants corresponding to these 2D planes using the polarization theory.

\section{Topological invariant from polarization theory (Zak phase)}\la{inv}
%%%%%%%%%%%%%%%%%%%%%%%%%%%%%%%%%%%%%%%%%%%%%%%%%%%%%%%%%%%%%%%%%%%%%%%%%%%%%%%%%%%%%%%%%%?
\begin{center}
\begin{figure}[htb!]
\textrm{(a)}\hspace{4cm}\textrm{(b)}
\includegraphics[width=4cm,height=2.5cm]{fig4a.pdf}\hspace{0.1cm}\includegraphics[width=4.0cm,height=2.5cm]{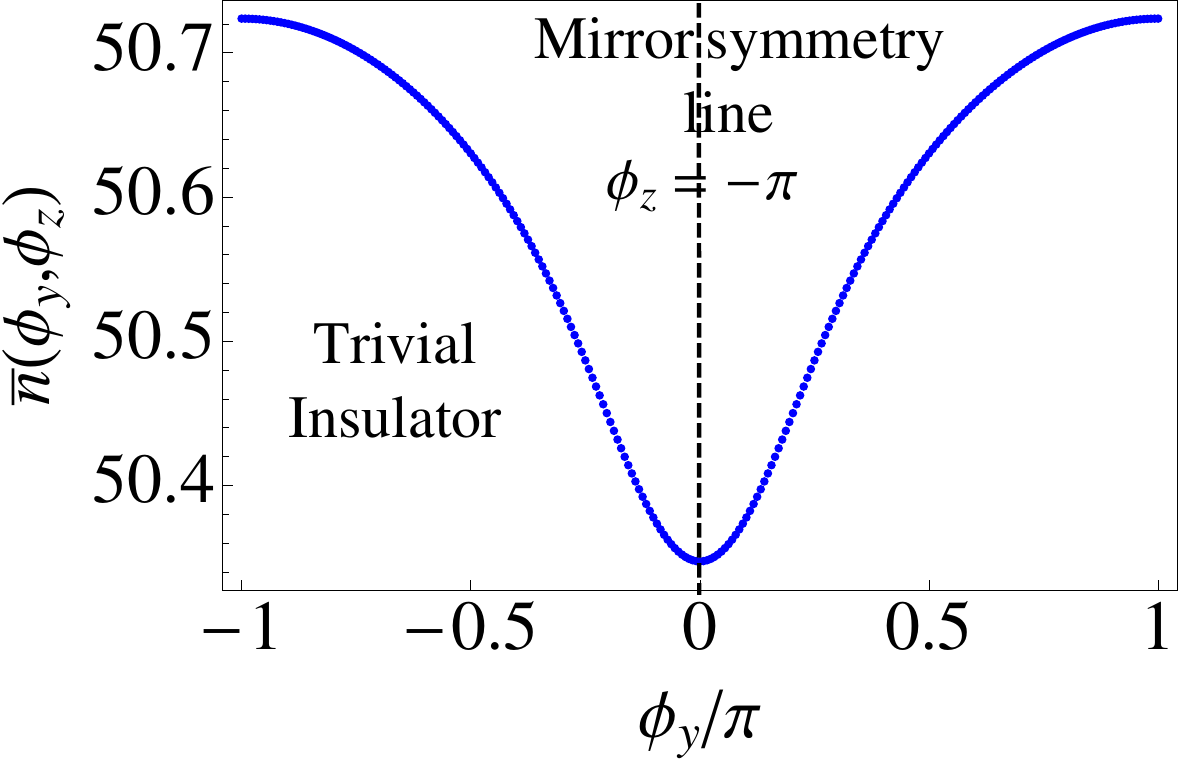}\\
\hspace{-0.2cm}\textrm{(c)}\hspace{4cm}\textrm{(d)}
\includegraphics[width=4.0cm,height=2.5cm]{fig4b.pdf}\hspace{0.1cm}\includegraphics[width=4.0cm,height=2.5cm]{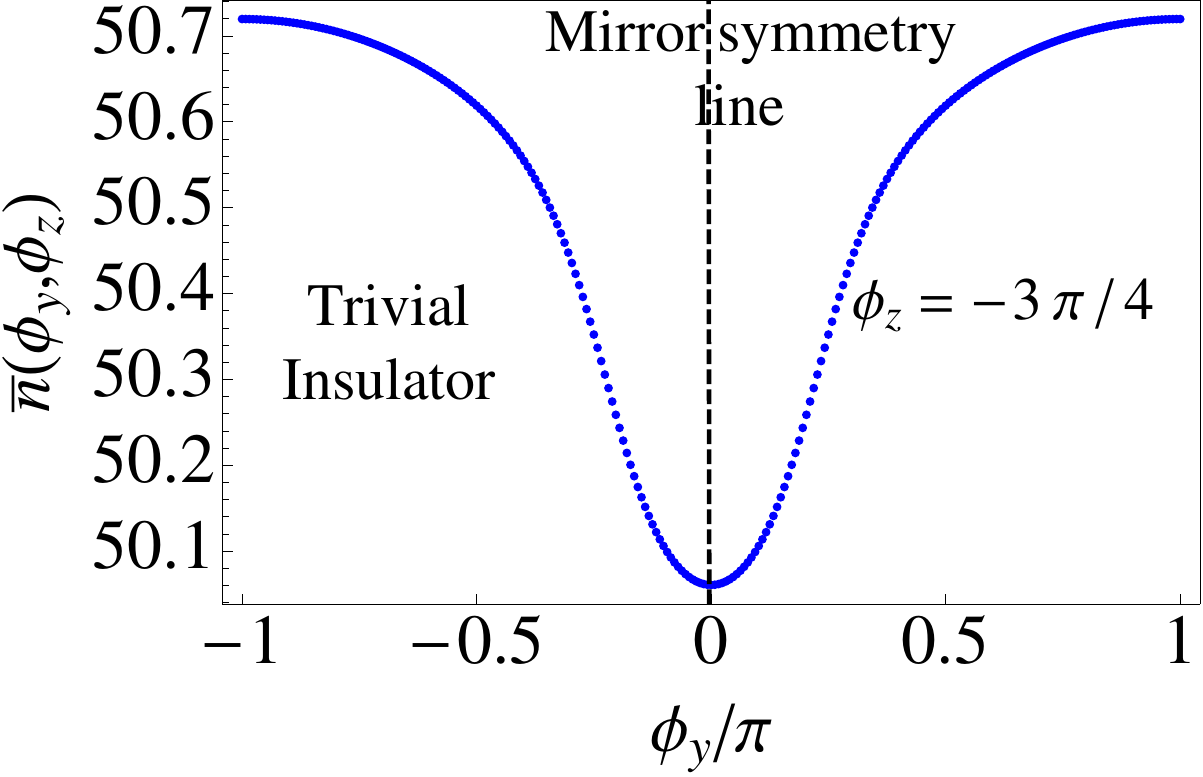}\\
\textrm{(e)}\hspace{4cm}\textrm{(f)}
\includegraphics[width=4cm,height=2.5cm]{fig4c.pdf}\hspace{0.1cm}\includegraphics[width=4.0cm,height=2.5cm]{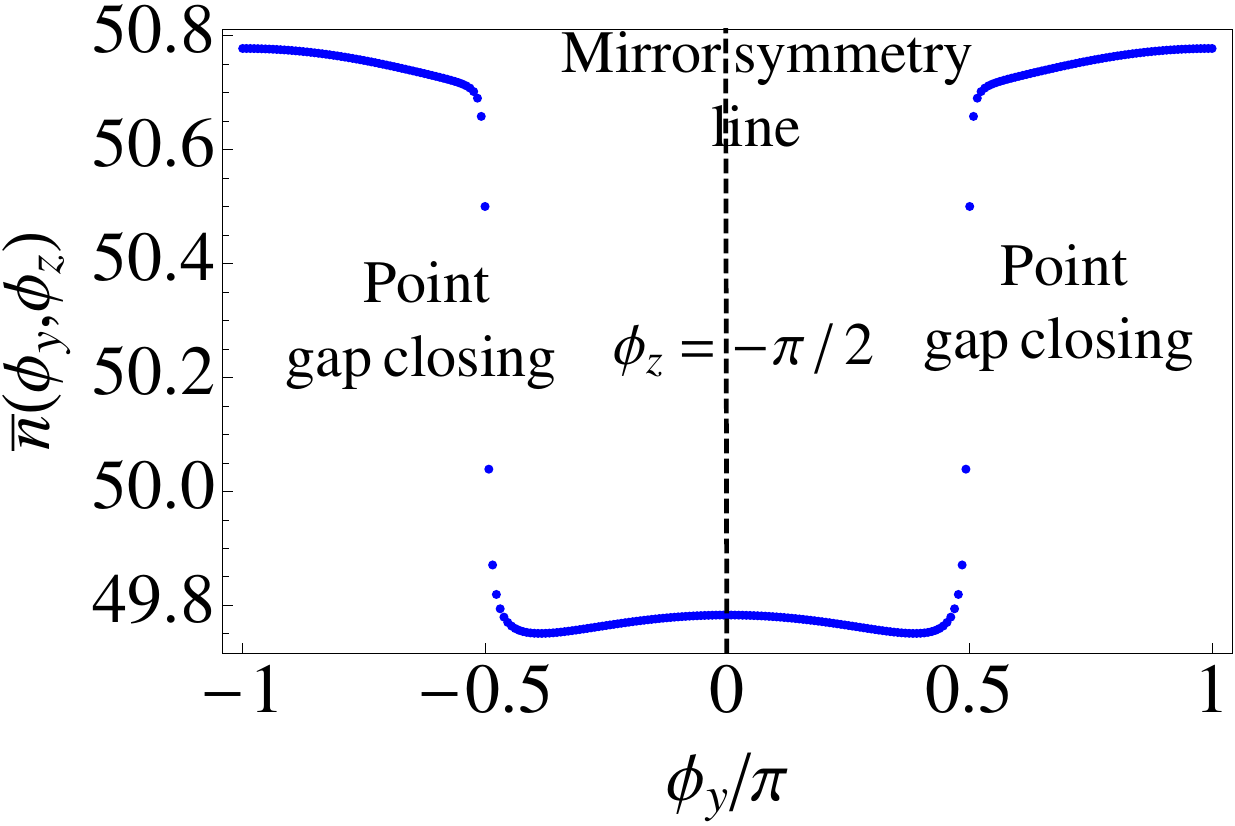}\\
\hspace{-0.2cm}\textrm{(g)}\hspace{4cm}\textrm{(h)}
\includegraphics[width=4.0cm,height=2.5cm]{fig4d.pdf}\hspace{0.1cm}\includegraphics[width=4.0cm,height=2.5cm]{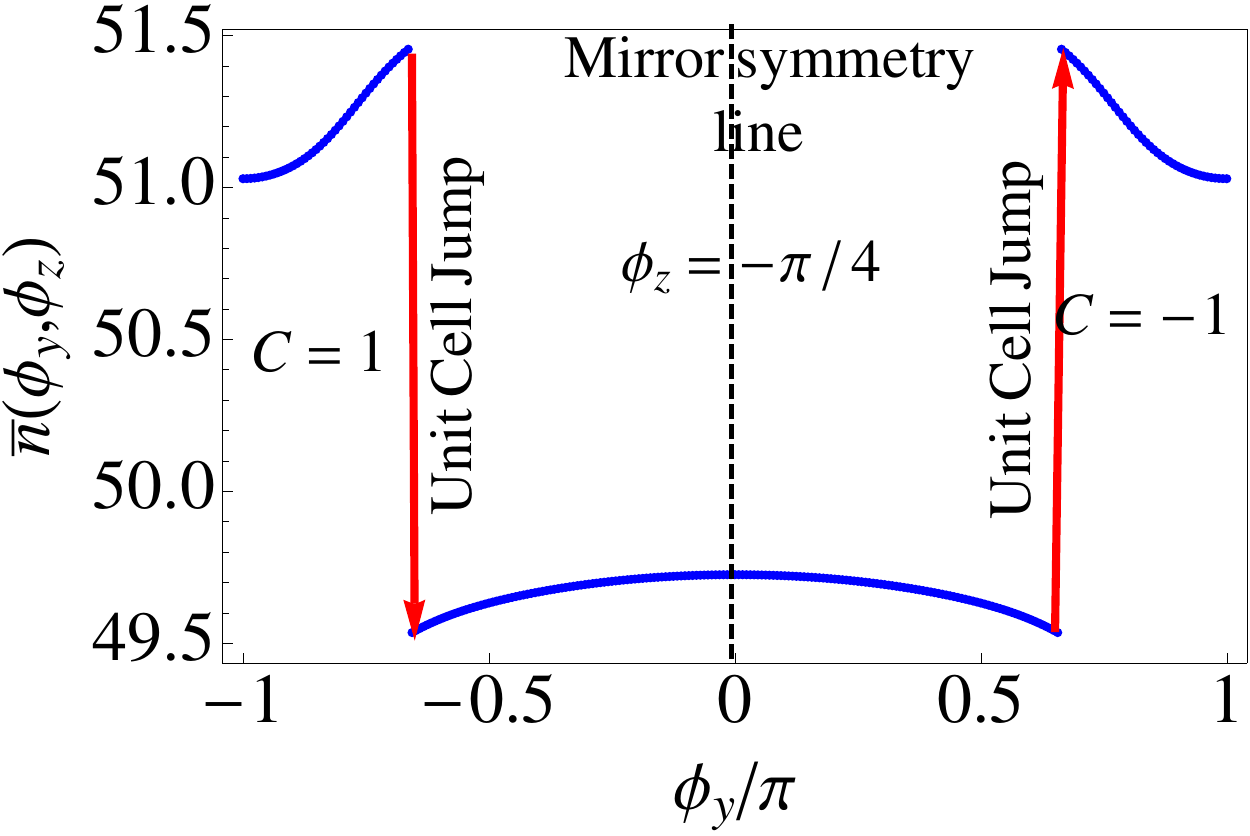}\\
\hspace{-0.2cm}\textrm{(i)}\hspace{4cm}\textrm{(j)}\\
\includegraphics[width=4.0cm,height=2.5cm]{fig4e.pdf}\hspace{0.1cm}\includegraphics[width=4.0cm,height=2.5cm]{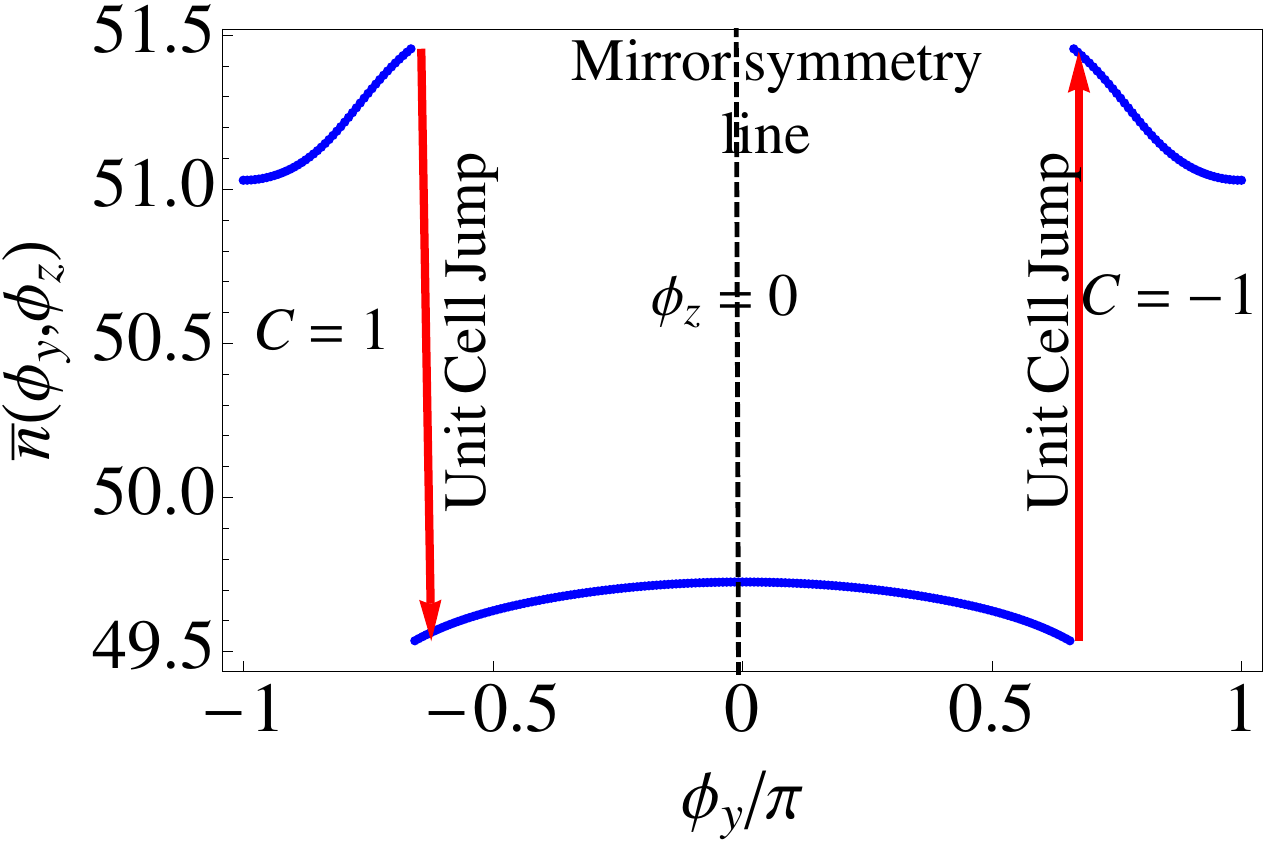}
\caption{ We plot average Hybrid wannier charge center for tight binding chain of 100 sites with open boundary conditions and parameter value $\lambda_{z}=\lambda_{xy}=0.4$,  $\lambda_{xz}=0.6,\ \lambda_{y}=0.4$, $t=1$. Plots  (b) and  (d)  show $\bar{n}(\phi_y,\phi_z) \, \text{vs}\  \phi_y$ for  trivial (insulator) brillouin zone slices $\phi_z=-\pi,\  -3\pi/4$.  Plots  (h), (j) show $\bar{n}(\phi_y,\phi_z) \, \text{vs}\  \phi_y$ for topologically non-trivial ( two mirror copies of the Chern insulator) BZ slices $\phi_z=-\pi/4,    \text{and}\ 0$.  Plot (f) show the topological transition at $\phi_z=-\pi/2$ which is a semimetallic 2D plane containing the band touching points or Weyl nodes. We plot the bands corresponding to the polarization plot for each 2D plane in Figs. (a, c, e, g, i). }
\label{pol}
\end{figure}
\end{center}
%%%%%%%%%%%%%%%%%%%%%%%%%%%%%%%%%%%%%%%%%%%%%%%%%%%%%%%%%%%%%%%%%%%%%%%%%%%%%%%%%%%%%%%%% In 
Polarization (P) or equivalently the Zak phase provides a natural framework to map the topological invariants\cite{vanderbilt,restaprb} within the AAH framework and has been recently put forward as a tool to study topological invariants in cold atoms\cite{troyer}. Modern theory of polarization of insulators\cite{vanderbilt,restaprb} is directly related to the geometry of underlying Bloch bands. 
\be
P (\text{Zak phase})=i\oint\langle u_k|\partial_k|u_k\rangle dk
\la{polg}
\ee
P on its own is a gauge non-invariant quantity and is defined modulo the choice of a unit cell. However, the change in polarization is a well defined (and gauge invariant) physical observable which captures the topological invariants for the underlying topologically non-trivial bands. Polarization is also sometimes expressed as a Zak phase which has been experimentally measures for an SSH model in a double well optical lattice\cite{bloch}. The polarization of a \emph{finite 1D insulator} can be expressed in terms of the average position of the charge centers of Hybrid Wannier Functions (HWF).
This polarization is a function of the phase parameters $(\phi_y,\phi_z)$ which act as an adiabatic flux under which the polarization changes. For topologically non-trivial systems of IQH type, there is a discontinuous change in the polarization that represents the topological charge transferred from one edge to the other edge, akin to the Laughlin IQHE gauge argument\cite{laughlin}. The average position of the charge center of HWF is given by\cite{troyer},
\begin{gather}
\bar{n}(\phi_y,\phi_z)=\frac{\sum_{n}\langle n \rho(n,\phi_y,\phi_z)\rangle}{\sum_{n}\langle\rho(n,\phi_y,\phi_z)\rangle},\nonumber\\
\rho(n,\phi_y,\phi_z)=\sum_{\text{occupied states}} |n,\phi_y,\phi_z\rangle\langle n,\phi_y,\phi_z|,
\la{polarization}
\end{gather}
where $n$ is the real space site index and $|n,\phi_y,\phi_z\rangle$ is the hybrid eigenstate of the system. 

In Fig.~(\ref{pol}), we plot the average HWF charge center as a function of $\phi_y$ for different slices of $\phi_z$. Figs.~(\ref{pol} b and d) correspond to the topologically trivial insulators where $\bar{n}$ changes continuously as a function of $\phi_y$. The topologically trivial insulating 2D slices of the Brillouin zone (see Figs.~\ref{pol} a and c) do not  have the band touching points and they have a smoothly varying Wannier charge center as a function of the phase $\phi_y$ as shown in Figs.~\ref{pol} b and d. The spectrum plots corresponding to the critical 2D planes (Fig.~\ref{pol}e) manifest the topological transition at $\phi_z=-\pi/2$ which is a semimetallic 2D plane containing the band touching points or Weyl nodes. The corresponding polarization plot for this critical plane with the Weyl nodes is given in Fig.~\ref{pol}f. The energy spectrum corresponding to topologically non-trivial 2D slices are given in Figs.~\ref{pol}g and i for which the HWF charge center has discontinuous jumps as shown in Figs.~\ref{pol} h and j  at the band touching points for the Brillouin zone slices $\phi_z=-\pi/4, 0$. Note that the discontinuous jumps always occur in pairs and correspond to the Chern number of opposite signs. The overall Chern number in $\phi_y$ plane must add to zero due to the presence of time reversal symmetry. Here we have normalized the size of the jump to the size of the unit cell and in proper units this is exactly the $\pi$ shift in the Zak phase observed in Ref.~(\onlinecite{bloch}). Note that the HWF charge center shows a jump of one unit cell for $-\pi/2<\phi_z<\pi/2$ which corresponds to the band touching points carrying Chern numbers of $\pm1$ similar to the case of the mirror image copies Chern insulators ($Z_2$ topological insulator) in the full Brillouin zone due to the presence of time reversal symmetry.

 Experimentally,  Zak phase imaging can provide a striking distinction between trivial  and non-trivial topological slices of $\phi_z$. For a trivial case there will be a continuous change in the Zak phase as a function of the parameter $\phi_y$ (see Figs.~\ref{pol}b, \ref{pol}d) whereas for a non-trivial case, there will be a sudden discontinuous change in the Zak phase at the band touching points as a function of the phase $\phi_y$ as shown in Figs.~\ref{pol} h and j. Changing $\phi_y$ amounts to changing the lattice configuration, which has been accomplished experimentally by applying a laser pulse instantaneously\cite{bloch}. 
 
 \section{Double well optical lattices as momentum space building blocks of topological semimetals} \la{ex}
 %%%%%%%%%%%%%%%%%%%%%%%%%%%%%%%%%%%%%%%%%%%%%%%%%%%%%%%%%%%%%%%%%%%%%%%%%%%%%%%%%%%%%%%%%?
\begin{center}
\begin{figure}[htb!]
\vspace{0.2cm}
\includegraphics[width=7cm,height=5cm]{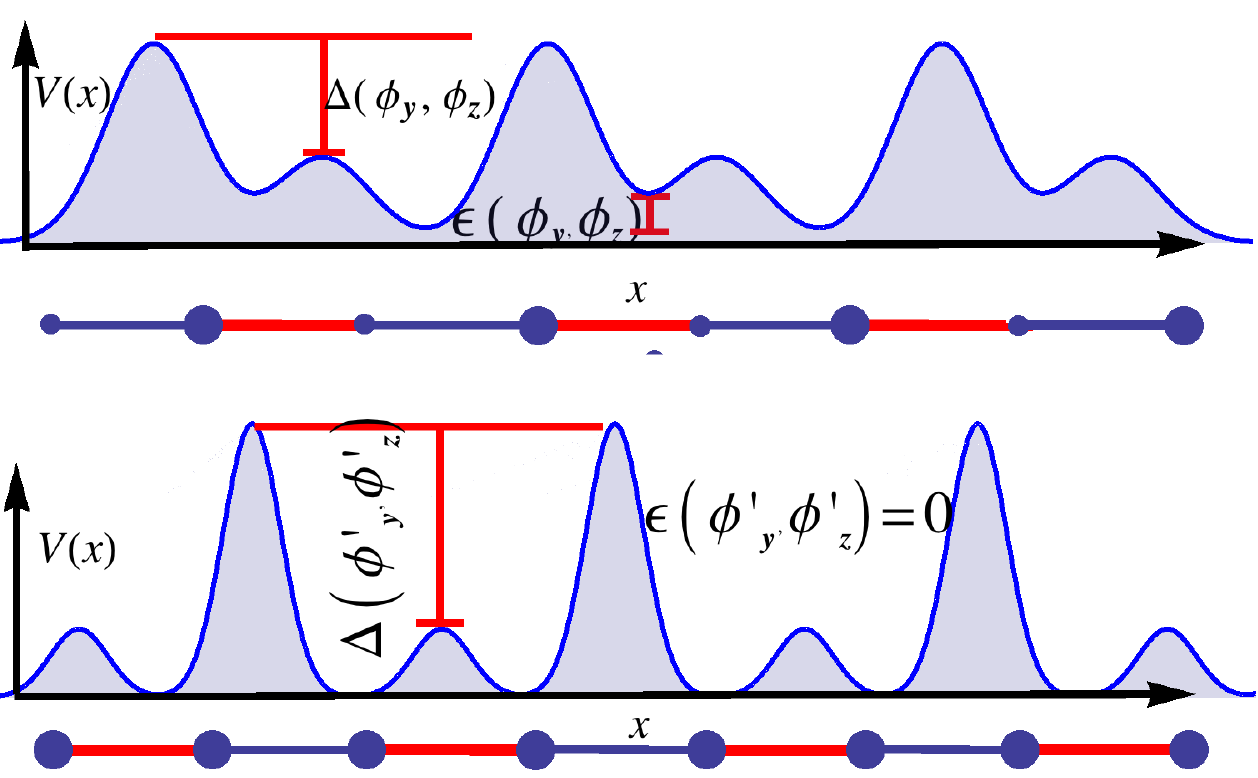}
\caption{Schematic of a double well potential with $\Delta(\phi_y,\phi_z)$ modulating the relative hopping amplitude of the dimerized state. $\epsilon(\phi_y,\phi_z)$ modulates the relative onsite energy of the different realizations of the 1D lattice.} \label{exp}
\end{figure}
\end{center}
%%%%%%%%%%%%%%%%%%%%%%%%%%%%%%%%%%%%%%%%%%%%%%%%%%%%%%%%%%%%%%%%%%%%%%%%%%%%%%%%%%%%%%%%%
One of our main motivations to understand 3D TS state using the generalized AAH framework is to make contact with the existing experimental technology in cold atoms and optical lattices, where the higher dimensional system can be (and often is) designed by combining suitable lower-dimensional lattices appropriately in a 'lego-building' type architecture. In recent years, cold atom experimentalists have not only made remarkable progress in realizing precisely designed optical lattices but also have exercised tremendous control over the various system parameters. Cold atom systems have an advantage over condensed matter systems as shown in the recently demonstrated single site control over the optical potential\cite{weitenberg}. Single site control allows access to physical observables like geometric phases going beyond the standard transport measurements in condensed matter systems and provides a new tool for understanding topological properties.
 Recently, a lot of progress has been made in experimentally simulating double-well potentials\cite{porto06, porto07,folling07,billy08,roati08}. The double well optical lattice (e.g. SSH model) has been used as a platform to directly image topological invariants of the Bloch bands by imaging the Zak phase\cite{bloch} which is directly related to the polarization of the 1D insulator given in Eq.~(\ref{polg}). In this experiment\cite{bloch}, different $\pi$-flux configurations (different $\phi_{(y)z}$) of the AAH (or the equivalent SSH) model were  generated by instantaneously applying laser pulses of different phases, which switch the lattice between different configurations.  The double well potential in terms of the experimental parameters can be written as,
 \bea
 V(x,\phi_y,\phi_z)&=&V_1 \sin^2(k x+\frac{\delta(\phi_y,\phi_z)}{2})\nonumber\\
 &+&V_2(\phi_y,\phi_z)\sin^2(2k x+\frac{\pi}{2})
\eea
This one-dimensional optical superlattice potential can be engineered experimentally\cite{folling07,bloch}  by superimposing two standing optical waves. In this experiment one can also control different lattice amplitudes that allow for relative detuning between the onsite modulations in the AAH framework. Note that in the most general 1D AAH model that we present in Eq.~(\ref{aah1dm}), for any fixed value of the phase $(\phi_y, \phi_z)$, the real space lattice corresponds to a particular configuration of a 1D double well lattice with an onsite energy and a relative detuning between the two barrier heights ($\Delta_{+}(\phi_y,\phi_z)-\Delta_{-}(\phi_y,\phi_z)$) as shown in Fig.~(\ref{exp}). The presence of an onsite modulation in the AAH model appears in the theory simply as a detuning between the relative depth of each lattice site ($\epsilon(\phi_y,\phi_z)$) (Fig.~(\ref{exp})).  Systematic tuning of both these parameters in experiments has been achieved in Ref.~(\onlinecite{bloch}). We believe that the 3D TS connected with the 3D Weyl systems discussed by us in the current work can be produced experimentally in optical lattices by realizing different configurations of double well lattices. Topological features can then be captured  experimentally in such a system by mapping out the Zak phase for each realization of the lattice configuration. 
  
 Thus the 3D version of the 2-parameter AAH model in Eq.~(\ref{aah1dm}) with two phase angles serves as an archetype to study the Brillouin zone of the TS phase in three dimensions which manifest fascinating features such as `Fermi arcs' and `nodal lines'. Experimentally the 3D BZ of the TS phase can be realized via double well optical lattices and their topological properties can be captured by mapping recently developed methods in imaging Zak phases (polarization change). 
 
 In addition to cold atom experiments, the AAH model can also be simulated in photonic waveguides as shown in Refs.~(\onlinecite{kraus1,verbin13,sriramaah}). In a waveguide setup, the direction of propagation of the injected light corresponds to the phase $\phi_{y }$. The off-diagonal AAH physics can be simulated by modulating the lattice spacings between the neighboring waveguides along the direction of propagation\cite{kraus1,verbin13}. The onsite modulation corresponds to the thickness or the refractive index of the wave guides. Both these aspects can be engineered to provide a photonic waveguide that can simulate the full three dimensional Brillouin zone of a TS.
  
\section{Conclusion} \la{conclude}
In this work, we derived a general one dimensional Aubry-Andre-Harper (AAH) model with two phase angles corresponding to momenta in other dimensions starting from the 3D cubic lattice model with a magnetic field in the $y-z$ plane. This generalized 2-parameter AAH model (see Eq.~(\ref{aah1dm})) has modulations in both hopping and onsite potential energies. These modulations are controlled by two phase angle parameters $(\phi_y, \phi_z)$. For a $\pi$ flux, the AAH model is simply a bichromatic 1D lattice with detuning in the onsite energies. In Sec.~(\ref{cssm}), we work with an inversion symmetric model without allowing for onsite terms. For this model we show both analytically and numerically that there are robust topological zero energy modes for a range of phase $\phi_y$ confined between the two band touching points. For different values of $\phi_z$, the separation between the band touching points containing the zero modes can be varied. Varying $\phi_z$ between $(0,2\pi)$ brings out the three dimensional character in a sense that the only way to get rid of the band touching points is to bring them together where the two band touching points collapse to a single point of zero energy. We also explicitly obtain the three dimensional zero energy Fermi surface containing these topological zero modes in the Brillouin zone spanned by $(\phi_y, \phi_z)$. In Sec.~(\ref{weyl}), we allow for the onsite modulations breaking inversion symmetry and show that there are ``topologically unavoidable" regions of $(\phi_y,\phi_z)$ for which there exist zero energy modes terminating at the band touching points. These zero energy modes are the manifestation of well known Fermi arcs or nodal lines that are the hallmark of three dimensional Weyl semimetals. We establish the topological properties of the 3D semimetallic phase in two ways. We analytically show that these band touching points come from the zero energy modes of the double Majorana chain which manifesting the $Z_2$ index. We numerically demonstrate discontinuous jumps in polarization (change in Zak phase) at these band touching points which is a signature of topological charge transfer from one end of the system to the other. Both these methods are natural frameworks to understand the topological properties of the 1D AAH model for different values of $(\phi_y, \phi_z)$.

Finally, in Sec.~(\ref{ex}), we show that the theory we develop can be practically implemented within the existing capabilities of experimentally realized double well lattices. Experimentalists have not only demonstrated amazing control over the lattice parameters such as the onsite modulations and the relative barrier heights but also have measured the topological invariants in the form of Zak phase. Thus exotic TS phases in three dimensions can be realized through the systematic implementations of different configurations of double well lattices as shown in Fig.~(\ref{exp}). Topological features of the full 3D Brillouin zone can be imaged by looking at the Zak phase for each realization of the double well lattice. We conclude by stating that we have developed a detailed theory as well as a practical protocol for how to build 3D (as well as 2D) TS systems using 1D double-well optical lattices as lego-type building blocks, including the WSM of great current interest. Our work connects several seemingly disparate concepts in condensed matter physics in one synergistic whole:  1D Aubry-Andre-Harper and Su-Shrieffer-Heeger models, Majorana chains and robust zero energy modes, Fermi arcs and nodal lines, WSM and related topological systems, topological quantum phase transitions, and Zak phase in the polarization theory.   The topological physics of the WSM arises from its one particle band structure and the associated symmetry properties.  In real solids, where the particles are electrons, interaction effects are invariably always present possibly masking (certainly complicating) the topological features.  In an optical lattice, however, one can work with dilute noninteracting bosons, thus bringing out purely the one-particle quantum physics, so the topological weyl physics may very well be better studied with bosons in optical lattices than with electrons in real solids. We believe that all the elements and building blocks for the laboratory realization of TS in optical lattices following our protocol already exist, and therefore, we hope that TS will soon be realized in cold atomic systems.

\section{Acknowledgements} SG acknowledges Xiaopeng Li for helpful discussions. This work is supported by JQI-NSF-PFC and ARO-Atomtronics-MURI. 
 
\appendix

\section{Two parameter AAH model from a cubic lattice model with a tilted magnetic flux.}\la{appendix}
%%%%%%%%%%%%%%%%%%%%%%%%%%%%%%%%%%%%%%%%%%%%%%%%%%%%%%%%%%%%%%%%%%%%%%%%%%%%%%%%%%%%%%%%%%?
\begin{center}
\begin{figure}[htb!]
\includegraphics[width=8.0cm,height=6cm]{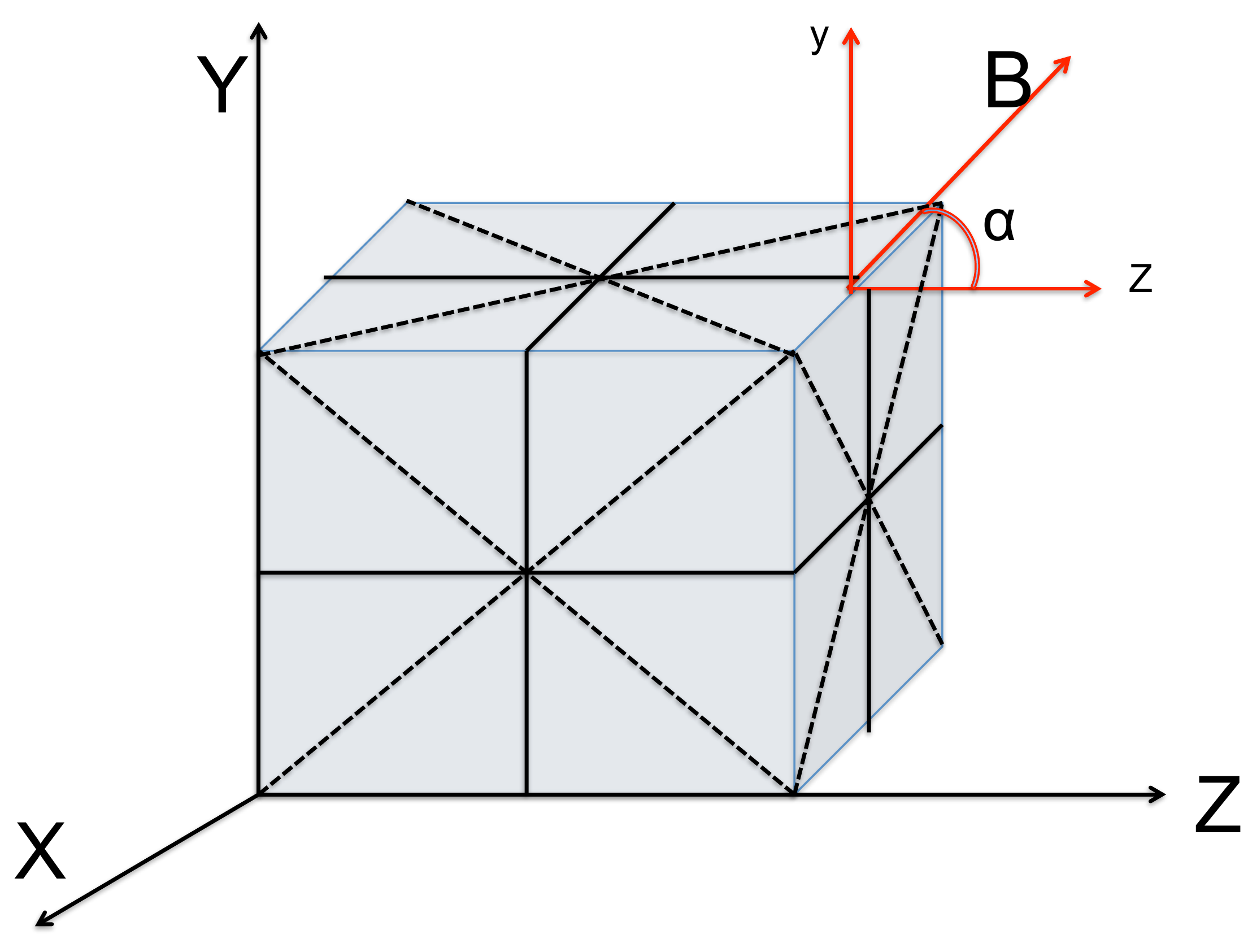}
\caption{Cubic lattice unit cell with tilted magnetic flux in y-z plane. The solid black lines denote the nearest neighbor hopping and the dashed black lines denote the off diagonal or next nearest neighbor hopping. B is the magnetic field confined to the y-z plane tilted by an angle $\alpha$ with respect to the z- axis.}
\label{cube}
\end{figure}
\end{center}
%%%%%%%%%%%%%%%%%%%%%%%%%%%%%%%%%%%%%%%%%%%%%%%%%%%%%%%%%%%%%%%%%%%%%%%%%%%%%%%%%%%%%%%%%
We start with the 3D tight binding Hamiltonian with nearest neighbor (NN) and next nearest neighbor (NNN) hopping  in the presence of a tilted magnetic field (See Fig.~(\ref{cube})). Such a real space model with a flux can be implemented with phase-engineered hopping in 3D optical lattices as shown in a recent proposal for WSM ~\cite{ketterle14}. We define $a$ as the lattice constant corresponding to the x, y and z directions spanning the cubic lattice. The AAH model we present in this paper can be derived starting from a 3D lattice model with a tilted flux. This connection allows us to study the topological features exclusive to 3D physics while working in the generalized 1D AAH framework. The Hamiltonian can be expressed as
\bea
H_{3D}&=& \sum_{n,m,l}\bigg[ t( e^{i \theta_{ij}} c^{\dagger}_{n+1,m,l}c_{n,m,l}+h.c.)\nonumber\\
          &+& \frac{\lambda_y}{2}( e^{i \theta_{ij}} c^{\dagger}_{n,m+1,l}c_{n,m,l}+h.c.)\nonumber\\
           &+& \frac{\lambda_z}{2}( e^{i \theta_{ij}} c^{\dagger}_{n,m,l+1}c_{n,m,l}+h.c.)\nonumber\\
           &+& \frac{\lambda_{xy}}{2}( e^{i \theta_{ij}} c^{\dagger}_{n+1,m+1,l}c_{n,m,l}+h.c.)\nonumber\\
           &+& \frac{\lambda_{xy}}{2}( e^{i \theta_{ij}} c^{\dagger}_{n-1,m+1,l}c_{n,m,l}+h.c.)\nonumber\\
           &+& \frac{\lambda_{xz}}{2}( e^{i \theta_{ij}} c^{\dagger}_{n+1,m,l+1}c_{n,m,l}+h.c.)\nonumber\\
           &+& \frac{\lambda_{xz}}{2}( e^{i \theta_{ij}} c^{\dagger}_{n-1,m,l+1}c_{n,m,l}+h.c.)\bigg],
\la{eq:Hamiltonianfull}
\eea
%\bea
%H_{3D}&=& t_{x}\sum_{<ij>_{x}} c^{\dagger}_{i}c_j e^{i \theta^{x}_{ij}}+\lambda^{y}_{ij}\sum_{<ij>_{y}}c^{\dagger}_{i}c_j e^{i \theta^{y}_{ij}}\nonumber\\
%&&+\lambda^{z}_{ij}\sum_{<ij>_{z}} c^{\dagger}_{i}c_j e^{i \theta^{z}_{ij}}.
%\la{eq:Hamiltonian}
%\eea
where $c$ and $c^{\dagger}$ are the usual Fermion creation and annihilation operators. $t$, $\lambda_{y\ (z)}$ and  $\lambda_{xy\ (xz)}$  correspond to the NN and NNN hopping  amplitude in the respective direction. $\theta_{ij}$  characterizes the magnetic field for the hopping between sites $i$ and $j$. It is convenient to express the magnetic field on the lattice in terms of flux per plaquette. Note that there is no flux through the $yz$ plane and the off-diagonal hopping in the $yz$ plane just rescales the energy and hence we can ignore this term without loss of generality. We can define the flux per plaquette as a line integral over the vector potential as,
\begin{gather}
\theta_{ij}=\frac{e}{hc}\int^{j}_{i} A.dl.
\end{gather}
We consider the magnetic field to lie in $yz$ plane with $B=(0,\ B\sin\alpha,\ B\cos\alpha)$, where $\alpha$ is the tilt angle with respect to the $z$-axis in the $yz$ plane. We choose a gauge corresponding to this magnetic field as $A=(0,\ Bx\cos\alpha,\ -Bx\sin\alpha)$. We define phases $\theta_{ij}$  on the links of the lattice for this gauge choice. We have defined $(b_y, b_z)=(b\cos\alpha,-b\sin\alpha)$. For a cubic lattice, we can define a flux per plaquette $b=\frac{Ba^2}{(hc/e)}$. The model involves both NN and NNN hopping therefore there are plaquettes of two different areas involved in the problem. The first one is the square plaquette enclosing $2\pi b_{y\ (z)}$  flux quantum in the $xy \ (xz)$ plane. The second one is the triangular plaquette which encloses $\pi b_{y\ (z)}$  flux quantum in the $xy\  (xz)$ plane. 
 \be
 \theta_{ij}=\begin{cases}
0\ \  & i=(n,m,l),\ j=(n+1,m,l) \\
2\pi nb_{z} & i=(n,m,l),\ j=(n,m+1,l)\\
2\pi nb_{y} & i=(n,m,l),\ j=(n,m,l+1)\\
\pi b_{z}(2n\pm1)\ \  & i=(n,m,l),\ j=(n+1,m\pm1,l)\\
\pi b_{y}(2n\pm1) & i=(n,m,l),\ j=(n+1,m,l\pm1)
\end{cases}
\la{flux}
\ee
We perform dimensional reduction of the 3D Hamiltonian to an effective 1D  tight binding model with phase parameters corresponding to the momenta in y and z directions. Using the following Fourier representation of the Fermion operators 
\be
c_{n,m,l}=\sum_{\phi_y,\phi_z}e^{i m \phi_y+i l \phi_z }c_{n}(\phi_y,\phi_z),\ \ c_{n}(\phi_y,\phi_z)\equiv c_n,
\la{fourier}
\ee
we can define the effective 1D Hamiltonian as,
\be
H_{3D}=\iint d\phi_y d\phi_z H_{1D}(\phi_y,\phi_z).
\la{3d1d}
\ee 
Using Eqs.~(\ref{flux}, \ref{fourier}, \ref{3d1d}) in the original 3D lattice Hamiltonian (Eq.~(\ref{eq:Hamiltonianfull})), we write the 1D AAH model with two phase parameters as,
\begin{eqnarray}
H_{1D}(\phi_{y},\phi_{z})&	=&\sum^{N-1}_{n=1}\bigg[\bigg\{t(1+\lambda_{xy}\cos(\pi(2n+1)b_{z}+\phi_{y})\nonumber\\
&+&\lambda_{xz}\cos(\pi(2n+1)b_{y}+\phi_{z})\bigg\}c_{n+1}^{\dagger}c_{n}+H.c.\nonumber\\
		&+&	\sum^{N}_{n=1}\bigg\{\lambda_{y}\cos(2\pi n b_{z}+\phi_{y})\nonumber\\
		&+&\lambda_{z}\cos(2\pi nb_{y}+\phi_{z})\bigg\} c_{n}^{\dagger}c_{n}\bigg].
		\la{1d}
  \end{eqnarray}
Starting from a 3D model, the phases $\phi_{y}$ and $\phi_z$ in diagonal and off-diagonal terms are shifted by a factor of $\pi b$ due to the two different plaquettes present in the system. Experimentally, one can design setups where the phases $\phi_{y,z}$ can be tuned independently, so we keep our notations general with  $\phi^{d}_{y,z}$ and  $\phi^{od}_{y,z}$ as independent variables for the diagonal and the off-diagonal definitions of the phases. A  simpler version of this 1D AAH model with just two onsite cosine terms has earlier been considered in the context of 3D IQHE\cite{kohmoto3d,halperin3d,aoki3d}.  The effective 1D AAH model can now be written in the following compact form (we drop the suffix 1D from the Hamiltonian),
 \begin{eqnarray}
&&H(\phi_y,\phi_z)=\nonumber\\
&&\sum^{N-1}_{n=1}t(1+\lambda_{xy}g_n(b_z,\phi^{od}_y)+\lambda_{xz}g_n(b_y,\phi^{od}_z))c_{n+1}^{\dagger}c_{n}+H.c.\nonumber\\
		&+&	\sum^{N}_{n=1}(\lambda_{y}g_n(b_z,\phi^{d}_y)+\lambda_{z}g_n(b_y,\phi^{d}_z)) c_{n}^{\dagger}c_{n}.
\la{aah1dapp}
\end{eqnarray}
We have defined $ g_n(b,\phi)=\cos(2n\pi b+\phi)$. 
%%%%%%%%%%%%%%%%%%
%%%%%%%%%%%%%%%%%
%%%%%%%%%%%%%%%%%%%%%%%%%%%%%%%%%%%%%%%
%%%%%%%%%%%%%%%%%%%%%%%%%%%%%%%%%%%%%%%
%%%%%%%%%%%%%%%%%%%%%%%%%%%%%%%%%%%%%%%
%\begin{thebibliography}{99}

\bibliographystyle{my-refs}

\bibliography{references.bib}

\end{document}